\documentclass[fleqn,usenatbib]{mnras}

\usepackage{newtxtext,newtxmath}

\usepackage[T1]{fontenc}

\DeclareRobustCommand{\VAN}[3]{#2}
\let\VANthebibliography\thebibliography
\def\thebibliography{\DeclareRobustCommand{\VAN}[3]{##3}\VANthebibliography}

\usepackage{graphicx}	
\usepackage{amsmath}	
\usepackage{xcolor}
\usepackage{tikz}
\usetikzlibrary{arrows.meta}

\newcommand{\code}[1]{\texttt{#1}}

\title[3D simulations of core helium burning]{3D hydrodynamics simulations of a 3\,$\mathbf{M_{\odot}}$ core-helium burning star}

\author[S. Blouin et al.]{
Simon Blouin$^{1,{\dagger}}$\thanks{E-mail: sblouin@uvic.ca},
Falk Herwig$^{1,{\dagger}}$,
Huaqing Mao$^{2,{\dagger}}$,
Pavel Denissenkov$^{1,{\dagger}}$,
and
Paul R. Woodward$^{2,{\dagger}}$
\\
$^{1}$Department of Physics and Astronomy, University of Victoria, Victoria, BC V8W 2Y2, Canada\\
$^{2}$LCSE and Department of Astronomy, University of Minnesota, Minneapolis, MN 55455, USA\\
$^{\dagger}$Joint Institute for Nuclear Astrophysics -- Center for the Evolution of the Elements (JINA--CEE)
}

\date{Accepted 2023 November 10}

\pubyear{2023}

\begin{document}
\label{firstpage}
\pagerange{\pageref{firstpage}--\pageref{lastpage}}
\maketitle
\begin{abstract}
The inner structure of core-helium burning (CHeB) stars remains uncertain due to the yet unknown nature of mixing at the boundary of their cores. Large convective cores beyond a bare Schwarzschild model are favoured both from theoretical arguments and from asteroseismological constraints. However, the exact nature of this extra mixing, and in particular the possible presence of semiconvective layers, is still debated. In this work, we approach this problem through a new avenue by performing the first full-sphere 3D hydrodynamics simulations of the interiors of CHeB stars. We use the \code{PPMstar} explicit gas dynamics code to simulate the inner 0.45$\,M_{\odot}$ of a 3\,$M_{\odot}$ CHeB star. Simulations are performed using different Cartesian grid resolutions (768$^3$, 1152$^3$ and 1728$^3$) and heating rates. We use two different initial states, one based on \code{MESA}'s predictive mixing scheme (which significantly extends the core beyond the Schwarzschild boundary) and one based on the convective premixing approach (which exhibits a semiconvective interface). The general behaviour of the flow in the convective core and in the stable envelope (where internal gravity waves are observed) is consistent with our recent simulations of core convection in massive main-sequence stars, and so are the various luminosity scaling relations. The semiconvective layers are dominated by strong internal gravity waves that do not produce measurable species mixing, but overshooting motions from the convective core gradually homogenize the semiconvective interface. This process can possibly completely erase the semiconvective layers, which would imply that CHeB stars do not harbour a semiconvection zone.
\end{abstract}


\section{Introduction}
\label{sec:intro}
Core helium burning (CHeB) stars are characterized by a central convective He-burning core surrounded by a convectively stable He-rich envelope. Observationally, CHeB stars are known as red clump stars, secondary clump stars, RR-Lyrae, horizontal branch stars, or subdwarfs B. These various classes of CHeB stars descend from different evolutionary pathways, but they all have a He-fusing core where the triple-$\alpha$ reaction produces carbon and $^{12}{\rm C}(\alpha,\gamma)^{16}{\rm O}$ makes oxygen.

\begin{figure}
    \centering
	\includegraphics[width=\columnwidth]{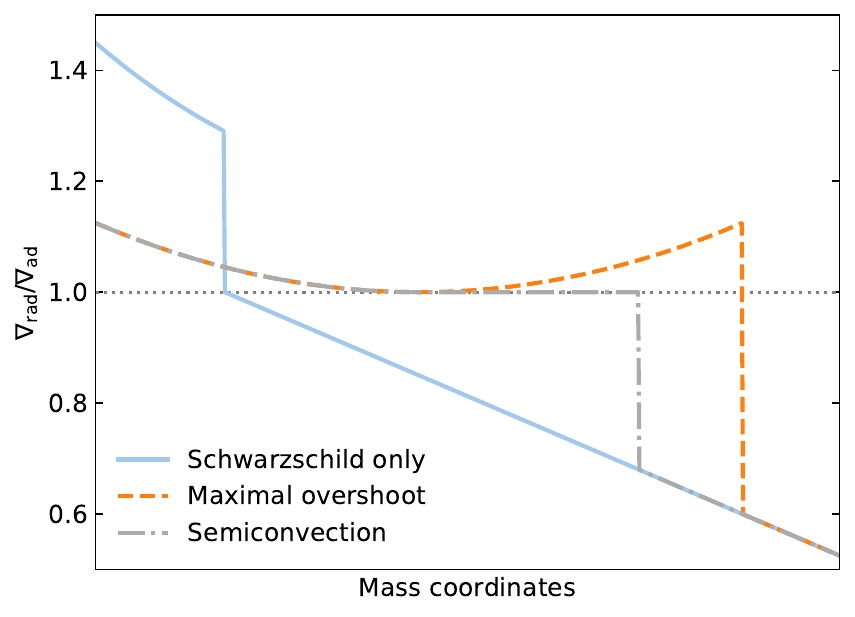}
    \caption{Qualitative representation of the radiative gradient profiles described in Section~\ref{sec:intro}. The actual temperature gradient $\nabla_{\rm T}$ is equal to $\nabla_{\rm ad}$ in the regions where $\nabla_{\rm rad}/\nabla_{\rm ad} > 1$ and equal to $\nabla_{\rm rad}$ otherwise.}
    \label{fig:nabla_cartoon}
\end{figure}

In low- and intermediate-mass stars, the treatment of convective boundary mixing (CBM) at the edge of the He-fusing core is particularly challenging for 1D stellar evolution \citep[e.g., see the review by][]{salaris2017}. The generation of C and O inside the convective core enhances its opacity $\kappa$, and the radiative gradient,
\begin{equation}
    \nabla_{\rm rad} = \frac{3}{16 \pi a c G} \frac{\kappa L P}{m T^4},
    \label{eq:nablarad}
\end{equation}
increases with time in the convection zone. In Equation~\eqref{eq:nablarad}, $a$ is the radiation constant, $c$ is the speed of light, $G$ is the gravitational constant, $L$ is the luminosity, $P$ is the pressure, $m$ is the mass enclosed within the radius at which $\nabla_{\rm rad}$ is calculated, and $T$ is the temperature. This $\nabla_{\rm rad}$ increase leads to the formation of a discontinuity, such that $\nabla_{\rm rad} > \nabla_{\rm ad}$ just inside the core and $\nabla_{\rm rad} = \nabla_{\rm ad}$ just outside (blue solid line in Figure~\ref{fig:nabla_cartoon}), where $\nabla_{\rm ad}$ is the adiabatic temperature gradient. By definition, the convective boundary is located at the radius where radiation carries out all the energy. Within the mixing length theory (MLT) framework used in many 1D stellar evolution calculations, this condition is satisfied where the Schwarzschild criterion $\nabla_{\rm rad} = \nabla_{\rm ad}$ is met \citep{biermann1932}. It is of course well established that convective penetration can significantly shift the location of the convective boundary with respect to the Schwarzschild boundary \citep[for recent multi-dimensional simulations demonstrating this point, see][]{kapyla2019,anders2022,andrassy2023,baraffe2023,blouin2023a,mao2023}, but it is instructive in what follows to consider the consequences of applying the standard Schwarzschild criterion to the He-burning core.

Within this framework, $\nabla_{\rm rad} = \nabla_{\rm ad}$ must be satisfied on the \textit{convective side} of the boundary, since otherwise the convective flux cannot be zero \citep{gabriel2014}. In CHeB stars, we encounter in 1D stellar evolution a situation where $\nabla_{\rm rad} = \nabla_{\rm ad}$ on the radiative side of the $\nabla_{\rm rad}$ discontinuity, but where $\nabla_{\rm rad} > \nabla_{\rm ad}$ on the convective side. Therefore, the convective flux is not zero at that location, and this cannot be the convective boundary. This conclusion can also be reached by realizing that the Schwarzschild boundary is in an unstable equilibrium \citep{schwarzschild1969,castellani1971a}. If the layer just above the convection zone is mixed with the core, its opacity increases due to the inflow of opaque C/O-rich material, $\nabla_{\rm rad}$ surpasses the adiabatic gradient, and the convective core grows. Mixing of the layer immediately above the core is inevitable: any amount of convective overshooting can accomplish that, and even atomic diffusion alone could be sufficient \citep{michaud2007}.

To find a stable boundary, the convective core must be extended until $\nabla_{\rm rad} = \nabla_{\rm ad}$ on the convective side of the Schwarzschild boundary, but this leads to further complications. Given the behaviour of the opacity and the thermodynamic structure of CHeB stars, extending the core leads to the formation of a local minimum in $\nabla_{\rm rad}$ within the convective region. Eventually, this local minimum in $\nabla_{\rm rad}$ reaches $\nabla_{\rm ad}$, effectively splitting the convective region in two. What does that mean for mixing in the star? Full mixing of the gap now separating the convective core from the convective shell above is problematic as it would imply the formation of a region with $\nabla_{\rm rad} < \nabla_{\rm ad}$ at the local minimum, in contradiction with the assumption of full mixing.

Two families of solutions have been proposed to solve this problem within the 1D MLT framework. The first consists of the formation of a partially mixed (or semiconvective) region between the $\nabla_{\rm rad}$ local minimum and the radiative envelope \citep{schwarzschild1969,paczynski1970,castellani1971b}. In this scenario, an extended semiconvective region where $\nabla_{\rm rad}=\nabla_{\rm ad}=\nabla_{\rm T}$ ($\nabla_{\rm T}$ is the actual temperature gradient) separates the convective core from the stable layers (grey dash-dotted line in Figure~\ref{fig:nabla_cartoon}). This CBM prescription has been implemented in many stellar evolution codes \citep[e.g.,][]{dorman1993,cassisi2003,constantino2015}, including in \code{MESA} through the convective premixing scheme \citep[CPM,][]{paxton2019,ostrowski2021}. Another solution consists of artificially halting the growth of the convective core before the splitting occurs. This is known as maximal overshoot \citep[][in the case where the growth is stopped just before the point where splitting would take place]{constantino2015,denissenkov2017,li2018} or predictive mixing \citep[PM,][]{paxton2018,ostrowski2021}.\footnote{The use of the term ``overshoot'' in this context can be confusing as the temperature gradient past the Schwarzschild boundary is affected, which contradicts a common definition of overshooting \citep[e.g.,][]{anders2023b}.} In this scenario, there is no partially mixed zone and, problematically, $\nabla_{\rm rad}$ remains greater than $\nabla_{\rm ad}$ on the inner side of the convective boundary (orange dashed line in Figure~\ref{fig:nabla_cartoon}).

It is unclear which of these two approaches should be preferred in 1D stellar evolution calculations, but some insights can be gained from observational constraints. Asteroseismological studies of various flavours of CHeB stars have clearly established that a bare Schwarzschild core (i.e., no extra CBM, as in the blue solid line of Figure~\ref{fig:nabla_cartoon}) can be ruled out. A more extended convective core is definitely required to reproduce the observed pulsation periods \citep{vangrootel2010b,vangrootel2010a,charpinet2011,charpinet2019,montalban2013,bossini2015,constantino2015,uzundag2021}. A similar conclusion is reached via cluster star counts, which are used to infer the lifetimes of horizontal branch stars \citep{constantino2016}. But beyond this finding, CHeB asteroseismological constraints do not yet offer a clear answer regarding which CBM scheme should be adopted. \cite{constantino2015,constantino2016} conclude that both the maximal overshoot and semiconvective prescriptions can be compatible with observations of horizontal branch stars. Similarly, for subdwarfs~B, \cite{uzundag2021} find that both the PM (or maximal overshoot) and CPM (or semiconvective) schemes implemented in \code{MESA} produce core masses that are in better agreement with the observations than bare Schwarzschild models (although even then the core sizes remain below the seismically derived values). Another promising observational window is white dwarf asteroseismology. The composition profiles of white dwarf C-O cores bear the imprint of the CHeB phase \citep{straniero2003,salaris2010,chidester2023}, and empirically derived white dwarf internal stratifications provide valuable constraints on CHeB CBM. \cite{giammichele2018,giammichele2022} have mapped the core compositions of a handful of white dwarfs and found large O-rich central regions that require extra CBM during the CHeB phase compared to standard evolution models \citep{degeronimo2019}.

Better understanding CBM in CHeB stars would have important ramifications beyond CHeB stars. In particular, extra mixing during the CHeB phase ultimately leads to the formation of white dwarfs with more O-rich cores, and the exact O abundance profile of white dwarfs is a key determinant of their cooling rates. Not only does it determine the total thermal energy content of the star \citep{fontaine2001}, but it also controls how much energy is released by fractionation processes during core crystallization \citep{segretain1994,montgomery1999,salaris1997,salaris2000,althaus2012,blouin2020,blouin2021}. Due to current uncertainties related to CBM during the CHeB phase, the O abundance profile of white dwarfs remains poorly constrained \citep{salaris2010,salaris2022}, and this injects systematic errors in the  white dwarf cooling models \citep{renedo2010,bedard2020,salaris2022,bauer2023} used in diverse age-dating applications \citep{hansen2013,fantin2019,boylan2021,kaiser2021,cimatti2023}. For the oldest white dwarfs in the Milky Way, CHeB CBM uncertainties result in errors of the order of a Gyr on inferred ages.

In this work, we use full-sphere 3D stellar hydrodynamics simulations of the interiors of CHeB stars to confront the CHeB CBM problem from a yet unexplored angle. Our simulations, performed with the \code{PPMstar} explicit gas dynamics code, follow for dozens of convective turnover timescales the 3D hydrodynamics response of the gas to a given CHeB structure. We study the behaviour of the flow for the two commonly used 1D CBM prescriptions, \code{MESA}'s PM and CPM schemes. We describe these \code{MESA} models in Section~\ref{sec:methods}, where we also explain how our \code{PPMstar} simulations are initialized. The general properties of the 3D simulations (flow morphology, convergence with respect to grid resolution, scaling relations) are explored in Section~\ref{sec:results}. In Section~\ref{sec:boundary}, we study mixing and entrainment in our simulations and discuss the implications of our findings in the context of the CHeB CBM problem. Finally, we conclude in Section~\ref{sec:conclu}.

\section{Methods}
\label{sec:methods}
In this section, we first describe the calculations we have performed with the \code{MESA} code to initialize our \code{PPMstar} simulations. In Section~\ref{sec:mapping} we then explain how the \code{MESA} models were mapped into our 3D simulations. Finally, we detail the \code{PPMstar} simulations themselves in Section~\ref{sec:ppmstar}.

\subsection{\code{MESA} evolution sequences}
\label{sec:MESA}
We use \code{MESA} version 12115 \citep{paxton2011,paxton2013,paxton2015,paxton2018,paxton2019} to generate our CHeB setups. We calculated the evolution of a $3\,M_{\odot}$ star from the pre-main sequence to the end of the CHeB phase. We chose an initial He mass fraction $Y=0.27$ and an initial metallicity $[{\rm Fe/H}]=-0.3$ using the \cite{asplund2009} abundance ratios. The mixing length is fixed to $\ell = 2 H_P$ (where $H_P$ is the local pressure scale height), and an overshoot parameter $f_{\rm ov}=0.015$ was assumed for the pre-CHeB evolution. When the beginning of the CHeB phase is reached, we separate the calculation into two distinct sequences: one using the PM scheme \citep{paxton2018} and one using the CPM scheme \citep{paxton2019}. For the PM calculation, we use a value of 0.005 for the \code{predictive\_superad\_thresh} parameter that controls the maximum extent of the convective core by enforcing a minimal value to $\nabla_{\rm rad}/\nabla_{\rm ad}-1$ in the mixed region. In the case of the CPM calculation, a very high spatial and temporal resolution is required to produce a smooth stratification at the core boundary: we use \code{mesh\_delta\_coeff=0.2} and \code{max\_years\_for\_timestep=10000}. The \code{MESA} inlists are available at \url{https://www.ppmstar.org}. We have also verified that using the more recent \code{MESA r22.11.1} for the CHeB evolution does not change the final stratification.

\begin{figure}
    \centering
    \includegraphics[width=\columnwidth]{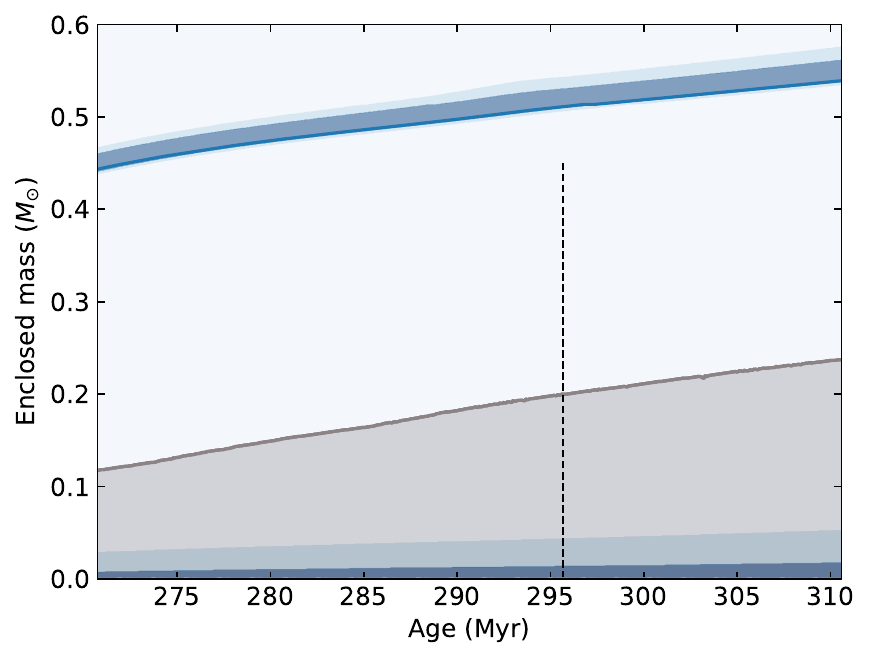}
    \caption{Kippenhahn diagram of the CHeB phase as calculated in our PM \code{MESA} run. The dark blue regions show where nuclear burning takes place (He burning at the center and H burning further out), and the grey region shows the extent of the convective core. The dashed vertical line marks the model used for our 3D \code{PPMstar} calculations; its vertical extent corresponds to the total mass included in the simulations.}
    \label{fig:kip}
\end{figure}

Figure~\ref{fig:kip} shows the Kippenhahn diagram of our PM \code{MESA} model during a portion of the CHeB phase. The convective core is in grey at the bottom and the nuclear burning regions are in dark blue (He burning at the center and H burning above in the stable envelope). The vertical dashed line indicates the particular model that we have elected to use as an initial state for our 3D simulations. It corresponds to a point where the central He mass fraction has reached $Y_{\rm c}=0.31$ and the He-burning luminosity is $L_{\star, {\rm He}}=62.5\,L_{\odot}$.\footnote{For the sake of brevity in our notation, we will denote the He-burning luminosity as $L_{\star}$ in what follows.} We have used the same value of $Y_{\rm c}$ to select the \code{MESA} model to use in the CPM case. Note that the vertical extent of the dashed line in Figure~\ref{fig:kip} represents the total mass included in our \code{PPMstar} simulations ($0<R<40\,$Mm in terms of radius): only a small portion of the envelope is considered and the H-burning shell is avoided. This is well justified here as anything that occurs above 40\,Mm is largely outside the scope of our study focused on mixing processes close to the convective core boundary. Extending the setup further out would decrease the grid resolution in the region of interest.

\subsection{\code{MESA}$\rightarrow$\code{PPMstar} mapping}
\label{sec:mapping}
\begin{figure}
    \centering
	\includegraphics[width=\columnwidth]{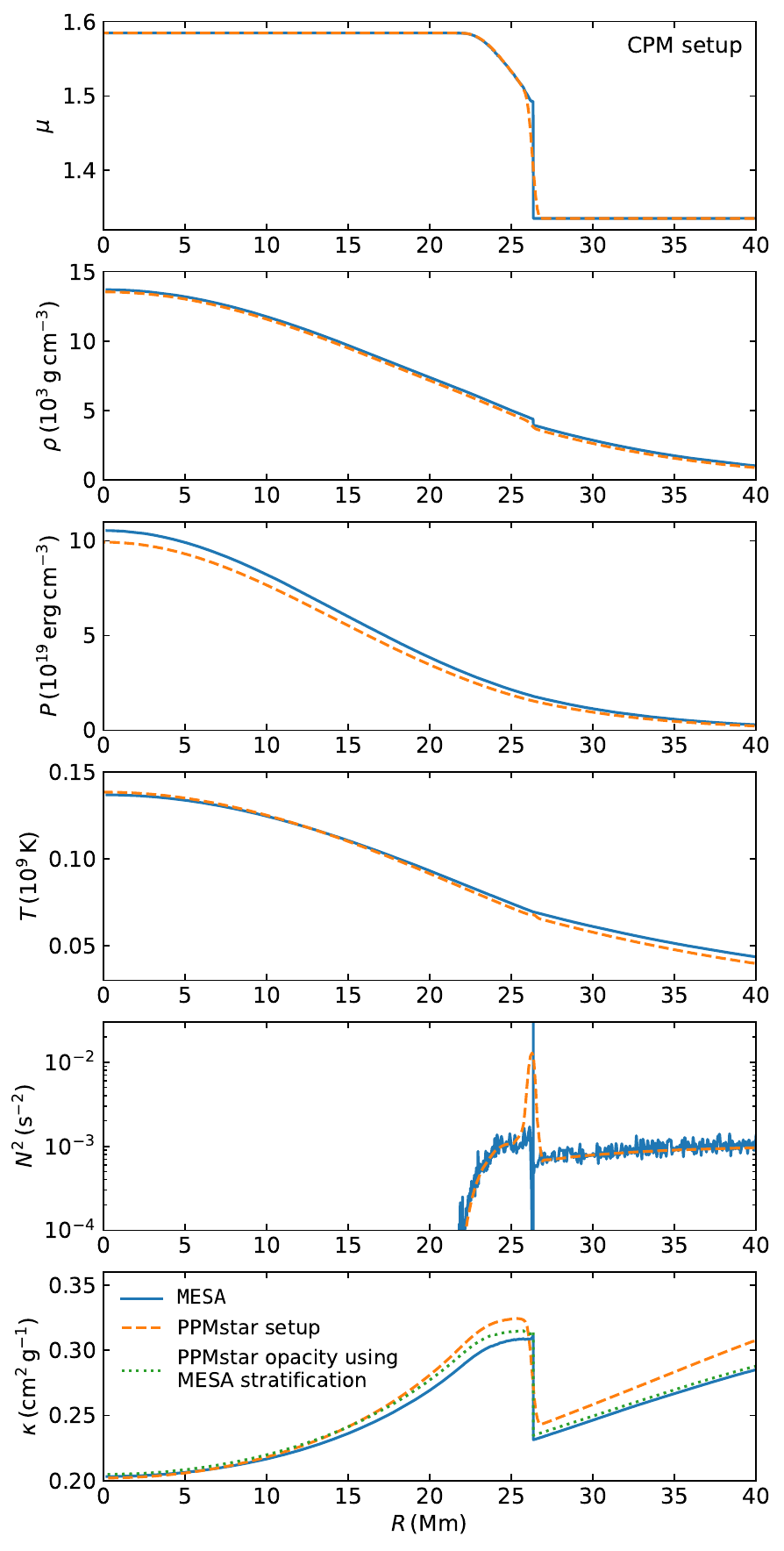}
    \caption{Radial profiles for our CPM setup. The \code{MESA} profile (blue solid line) and the base state of our \code{PPMstar} simulations (dashed orange line) are shown for each quantity. In the last panel, the dotted green line shows the opacity profile obtained using the \code{MESA} stratification and the \code{PPMstar} opacity module. The semiconvection zone is the region between $R \simeq 22$ and 26\,Mm characterized by a shallow $\mu$ gradient.}
    \label{fig:CPM_setup}
\end{figure}

\begin{figure}
    \centering
	\includegraphics[width=\columnwidth]{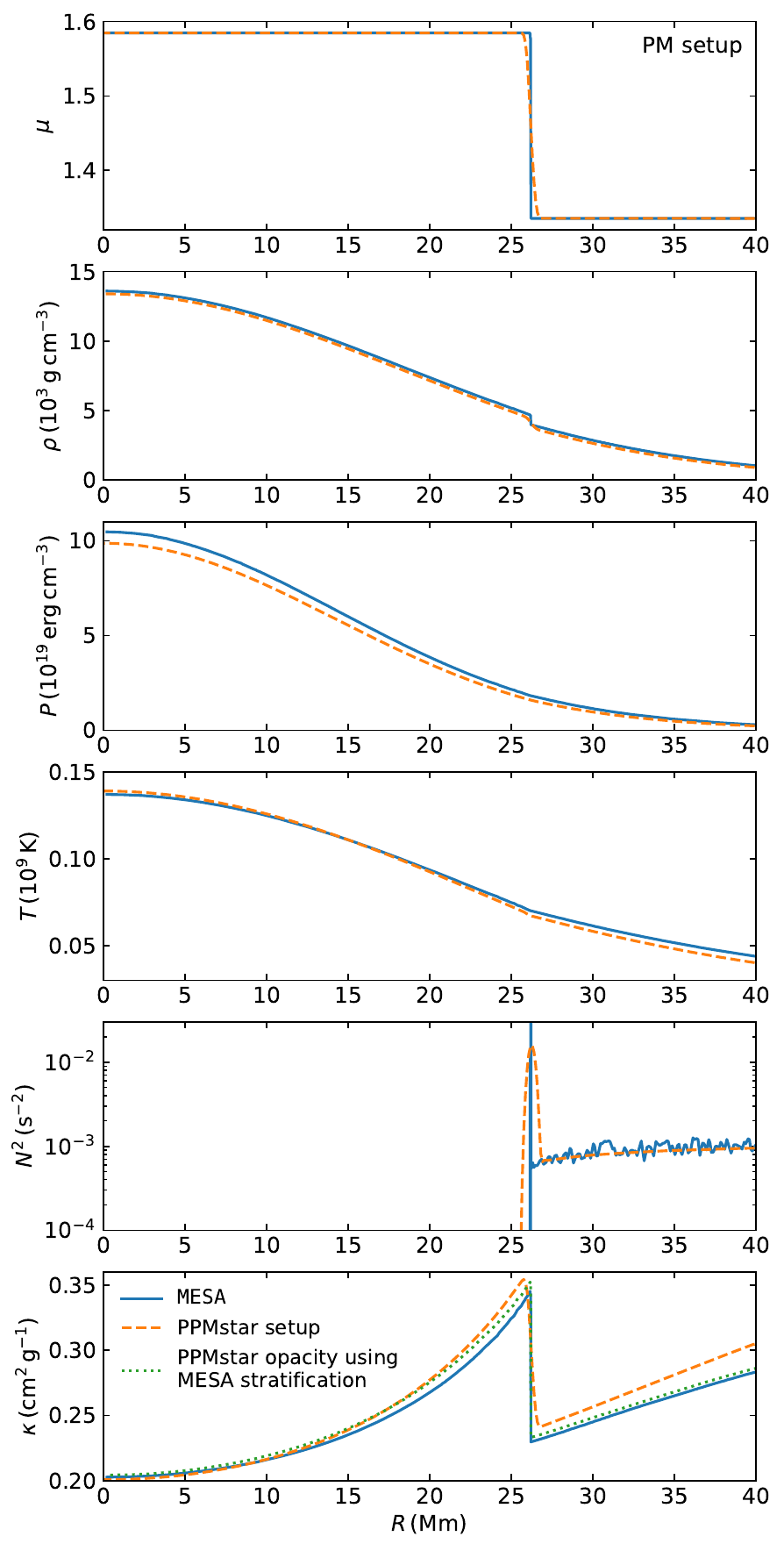}
    \caption{Same as Figure~\ref{fig:CPM_setup} but for our PM setup. }
    \label{fig:PM_setup}
\end{figure}

As in our recent \code{PPMstar} simulations of various types of stars \citep{blouin2023b,blouin2023a,herwig2023,mao2023}, the initial 3D state is reconstructed from the \code{MESA} entropy ($S$) and mean molecular weight ($\mu$) profiles. The \code{PPMstar} base state is calculated using these profiles and enforcing hydrostatic equilibrium. This integration is performed using \code{PPMstar}'s equation of state, which guarantees that our initial state is precisely in hydrostatic equilibrium. Note that small-scale noise was filtered out from the raw \code{MESA} $S$ and $\mu$ profiles to avoid injecting spurious small-scale structures into the 3D base state. In addition, we smoothed the $S$ and $\mu$ profiles at the outer edge of the semiconvection zone (in the CPM case) and of the convective core (in the PM case) as the CPM and PM schemes each produce unphysical discrete jumps. These transition regions were flattened so that they span at least 10 grid cells in our \code{PPMstar} simulations. This procedure is shown in the top panels of Figures~\ref{fig:CPM_setup} and~\ref{fig:PM_setup}.

The equation of state currently implemented in \code{PPMstar} includes the pressure contributions from the ideal gas and from the radiation field,
\begin{equation}
    P = P_{\rm gas} + P_{\rm rad} = \frac{R \rho T}{\mu} + \frac{a T^4}{3},
    \label{eq:eos}
\end{equation}
where $R$ is the ideal gas constant. In the central layers of CHeB stars, additional contributions from electron degeneracy pressure and ion--ion nonideal interactions come into play. As a result there is a $\simeq 10\%$ mismatch between the central pressure in our \code{MESA} models and that in our \code{PPMstar} setups (see third panels of Figures~\ref{fig:CPM_setup} and~\ref{fig:PM_setup}). A difference of that magnitude is not expected to impact the dynamics of the simulations in any meaningful way. Indeed, in our previous work on massive main-sequence stars, we have found the neglect of the radiation pressure term (which accounts for $\simeq 20\%$ of the pressure in the $25\,M_{\odot}$ star we have studied) to be an excellent approximation \citep{herwig2023,mao2023}. However, we will see in Section~\ref{sec:boundary} that this mismatch complicates the interpretation of the observed migration of the convective/semiconvective boundary over the course of our simulations.

Our \code{PPMstar} simulations include radiative diffusion via a radiative flux term in the energy equation \citep{mao2023},
\begin{equation}
    {\bf{F}}_{\rm rad} = - \frac{4 a c T^3}{3 \kappa \rho} {\bf \nabla} T.
\end{equation}
To model the opacity $\kappa$, we use the OPAL tables \citep{iglesias1996}. Since table look-ups would be too inefficient for a highly optimized code like \code{PPMstar}, we have built a polynomial expression that reliably approximates the OPAL tables in the limited composition--temperature--density space explored in our simulations (Appendix~\ref{sec:opacity}). As shown in the bottom panels of Figures~\ref{fig:CPM_setup} and~\ref{fig:PM_setup}, this simple prescription reliably captures the dependence of $\kappa$ on the composition and physical conditions. The green dotted lines obtained with this opacity model closely match the opacities calculated by \code{MESA} (blue solid lines). The opacity profile actually used in our simulations (orange dashed lines) departs more significantly from the \code{MESA} profiles. This is caused by the differences between the thermodynamic structures of the star in \code{MESA} and in \code{PPMstar} due to the incomplete equation of state.

\subsection{\code{PPMstar} simulations}
\label{sec:ppmstar}
\code{PPMstar} is an explicit gas dynamics code where the conservation equations are solved on a 3D Cartesian grid \citep{woodward2015,woodward2018,woodward2019,jones2017,andrassy2020,herwig2023}. In our simulations, the nuclear energy source from He burning in the core is modelled by a constant volume heating following a Gaussian radial profile that matches the shape of the \code{MESA} heating profile (our use of boost factors, described below, implies that we do not match its magnitude). Two fluids are included in the calculations: one having the mean molecular weight of the C/O-rich core ($\mu_{\rm core}=1.5845$) and one having the mean molecular weight of the almost pure-He envelope ($\mu_{\rm env}=1.3359$).

All our simulations are performed with heating luminosities that exceed the nominal He burning luminosity $L_{\star}$ of the star. MLT predicts a convective Mach number smaller than $10^{-4}$ in the He-burning cores of our $3\,M_{\sun}$ stars. As \code{PPMstar} is an explicit gas dynamics code, accurately resolving such slow flows would demand prohibitively small simulation grid cells. To circumvent this problem, we apply a boost factor to $L_{\star}$. We will present heating series (i.e., series of simulations that are identical except for their heating boost factors) in Section~\ref{sec:heating} that can be used to extrapolate our results to nominal luminosity. Another benefit of calculating heating series is that deviations from established scaling laws at low luminosities can be used to identify numerical resolution issues (e.g., when the flow becomes too slow for our explicit gas dynamics code; \citealt{herwig2023}). Note that the radiative conductivity,
\begin{equation}
    K = \frac{4 a c T^3}{3 \kappa \rho},
\end{equation}
is always multiplied by the same boost factor to ensure energy conservation in the star. If more heat is generated in the central layers, then it must be transported more efficiently by radiation. We also perform simulations for three different grid resolutions to assess the numerical convergence of our calculations (Section~\ref{sec:resolution}). All simulations discussed in this work are listed in Table~\ref{tab:runs}.

\begin{table}
  \centering
  \caption{Summary of simulations used in this work.}
  \label{tab:runs}
\begin{tabular}{cccccc}
\hline
 Run ID & Setup &  Grid & $\log L/L_{\star}$ &  \# dumps & Duration (h) \\
\hline
 W10 &   CPM &   768$^3$ &     6.0 &     2324 &         56.3 \\
 W11 &    PM &   768$^3$ &     6.0 &     1958 &         47.3 \\
 W12 &   CPM &   768$^3$ &     5.0 &     1080 &         26.1 \\
 W13 &    PM &   768$^3$ &     5.0 &     1227 &         29.6 \\
 W16 &   CPM &   768$^3$ &     4.5 &     1092 &         26.4 \\
 W17 &    PM &   768$^3$ &     4.5 &     1266 &         30.6 \\
 W20 &   CPM &  1728$^3$ &     5.0 &     1318 &         31.9 \\
 W21 &    PM &  1728$^3$ &     5.0 &     1348 &         32.6 \\
 W22 &   CPM &  1152$^3$ &     5.0 &     1718 &         60.6 \\
 W23 &    PM &  1152$^3$ &     5.0 &     1621 &         61.1 \\
 W24 &   CPM &   768$^3$ &     5.5 &     1097 &         26.6 \\
 W25 &    PM &   768$^3$ &     5.5 &     1329 &         32.1 \\
 W26 &   CPM &  1152$^3$ &     5.5 &     1193 &         48.1 \\
 W27 &    PM &  1152$^3$ &     5.5 &     1215 &         48.9 \\
 W28 &   CPM &  1152$^3$ &     6.0 &     1196 &         48.3 \\
 W29 &    PM &  1152$^3$ &     6.0 &     1039 &         41.8 \\
 W30 &   CPM &  1152$^3$ &     4.5 &     1179 &         47.6 \\
 W31 &    PM &  1152$^3$ &     4.5 &     1204 &         48.5 \\
 W32 &   CPM &   768$^3$ &     7.0 &     1706 &         41.3 \\
 W33 &    PM &   768$^3$ &     7.0 &     1513 &         36.6 \\
\hline
\end{tabular}
\end{table}

The analysis of our simulations relies on three types of outputs \citep{andrassy2020,stephens2021,herwig2023}. Every 1000--3000 time steps (depending on the grid resolution), a detailed output (``dump'') is written to disk. Each of these dumps contains spherically averaged profiles, high-precision 3D briquette data (on a grid that is four times smaller in each direction than the simulation grid), and full-resolution byte-sized data cubes that we use to generate qualitative visualizations of the flow.

\begin{figure*}
    \centering
	\includegraphics[width=\linewidth]{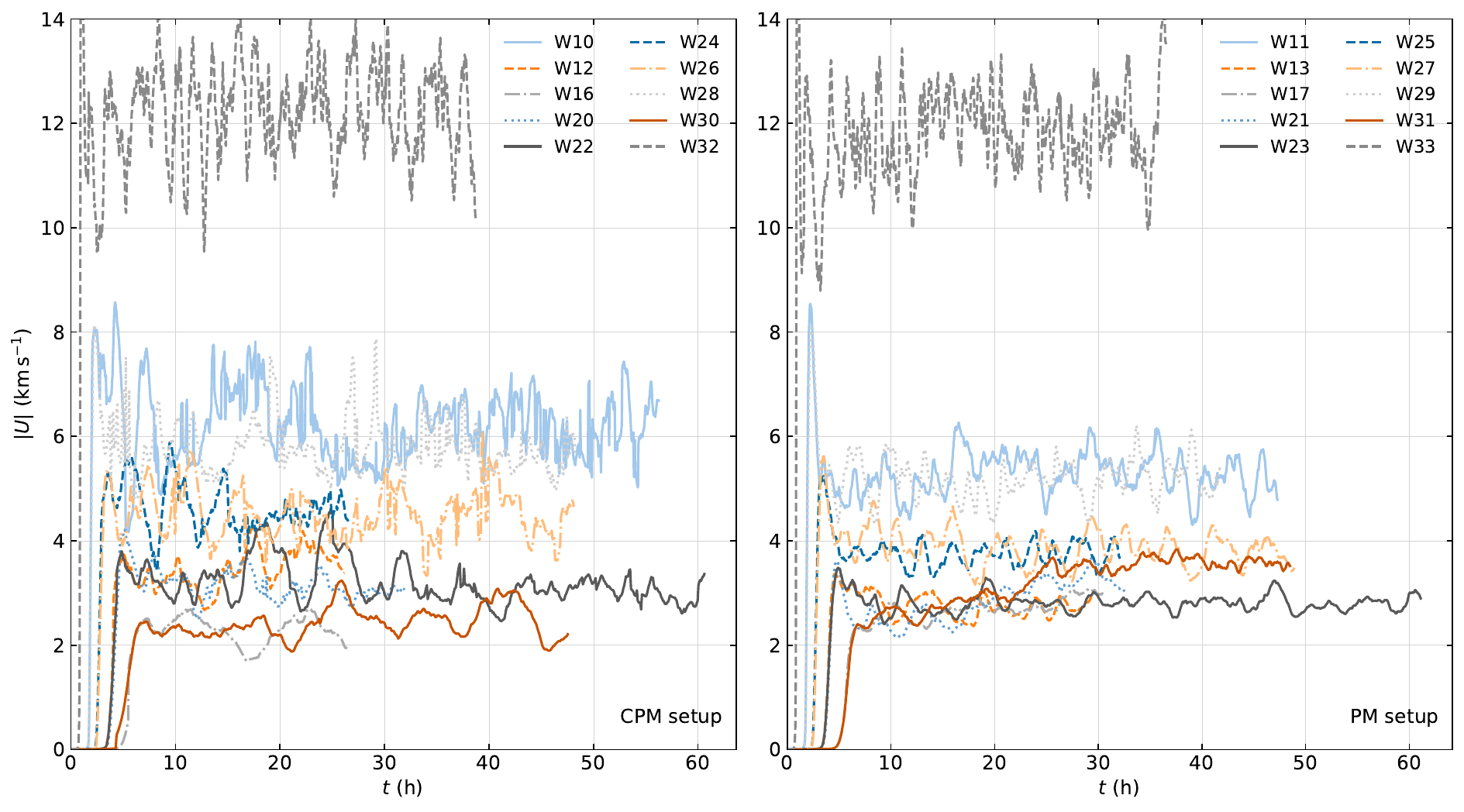}
    \caption{Time evolution of the spherically averaged rms convective velocity $|U|$ at a radius located $2\,H_P$ (15\,Mm) inside the convective boundary. The left panel shows the runs using the CPM setup and the right panel the PM setup. Most simulations reach a stationary state after a few convective turnover timescales.}
    \label{fig:timeseries}
\end{figure*}

As shown in Table~\ref{tab:runs}, each simulation is run for 25--60\,h of star time. As we will see in the next section, the rms velocity in the convective core for the $\log L/L_{\star}=5$ simulations is of the order of $U=4\,{\rm km}\,{\rm s}^{-1}$. This implies a convective turnover timescale of $\sim 2 \times 25\,{\rm Mm} / 4\,{\rm km}\,{\rm s}^{-1} \sim 3\,$h. Our simulations therefore span $\sim 10-50$ convective turnover timescales, depending on the heating rate \footnote{$U \propto L^{1/3}$ in the convective core (e.g.,  \citealt{porter2000,muller2016,jones2017,baraffe2021,herwig2023}).} and total simulation length. This is sufficient to robustly measure the mean properties of the flow. We show in Figure~\ref{fig:timeseries} the time evolution of the spherically averaged convective velocity at a radius located $2\,H_P$ (15\,Mm) inside the convective boundary. We can see that for almost all simulations, a state that appears to be stationary on the convective timescale is reached after just a few turnover timescales ($\sim 10\,$h). The initial transient before that time, when the convective flow is still building up, is discarded from our analysis in the rest of this work.

To quantify the absence of drifts in the velocity time series, we used linear regression to compute the slopes and estimate the percentage changes over the duration of each simulation. If we discard the first 10\,h of each run from the linear regression, we find that only four simulations (W12, W17, W21 and W31) exhibit a change of more than 15\% over their respective time durations. Most simulations are therefore unaffected by significant velocity drifts, and reliable time averages can be calculated. Note that two of the drifting simulations (W17 and W31) are low-heating runs at $\log L/L_{\star}=4.5$ that take more time to reach a stationary state due to the slower dynamics. As we will see, these low-heating simulations also suffer from numerical resolution issues (Section~\ref{sec:heating}). The results of these runs have to be interpreted with caution, and we will ignore them in most of our analysis.

\section{General properties of the simulations}
\label{sec:results}
\subsection{Velocity renderings}
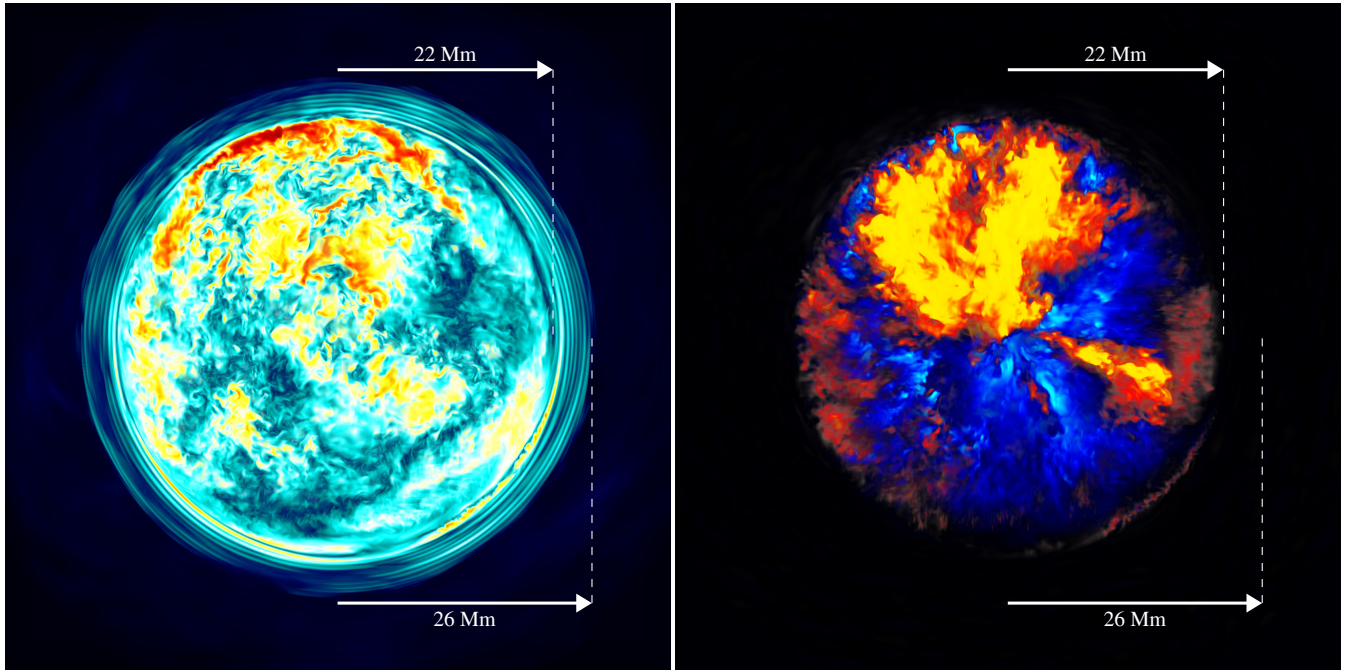
\begin{figure*}
    \centering
    \begin{tikzpicture}
        \node[anchor=south west,inner sep=0] (image) at (0,0) {\includegraphics[trim={7cm 7cm 7cm 7cm},clip,width=0.49\linewidth]{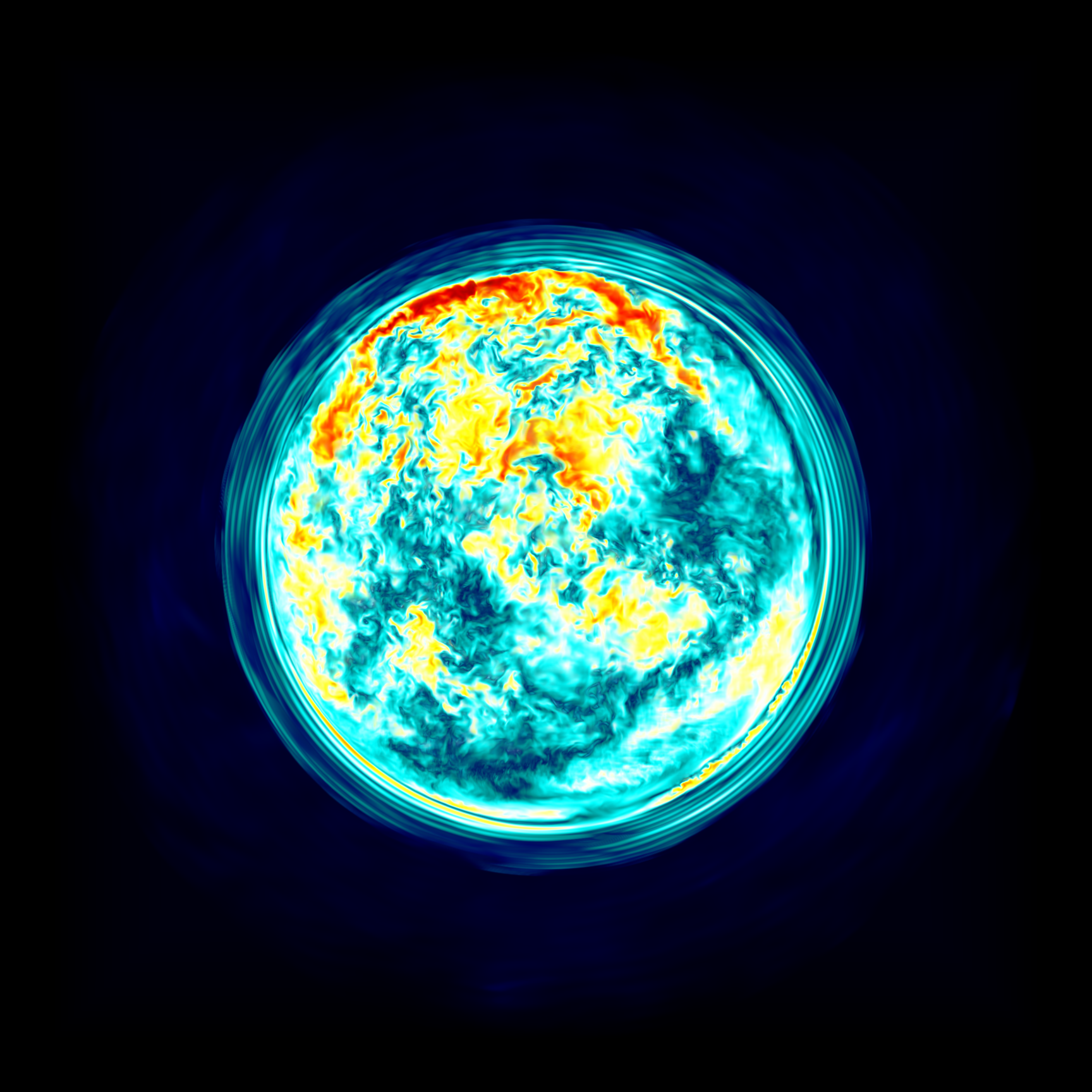}};
        \begin{scope}[x={(image.south east)},y={(image.north west)}]
            \draw[white,-Bar,-Triangle, line width=0.4mm] (0.5,0.9) -- (0.8235,0.9) node[midway,above] {22 Mm};
            \draw[white,dashed] (0.8235,0.9) -- (0.8235,0.5);
            \draw[white,-Bar,-Triangle, line width=0.4mm] (0.5,0.1) -- (0.882,0.1) node[midway,below] {26 Mm};
            \draw[white,dashed] (0.882,0.1) -- (0.882,0.5);
        \end{scope}
    \end{tikzpicture}
    \begin{tikzpicture}
        \node[anchor=south west,inner sep=0] (image) at (0,0) {\includegraphics[trim={7cm 7cm 7cm 7cm},clip,width=0.49\linewidth]{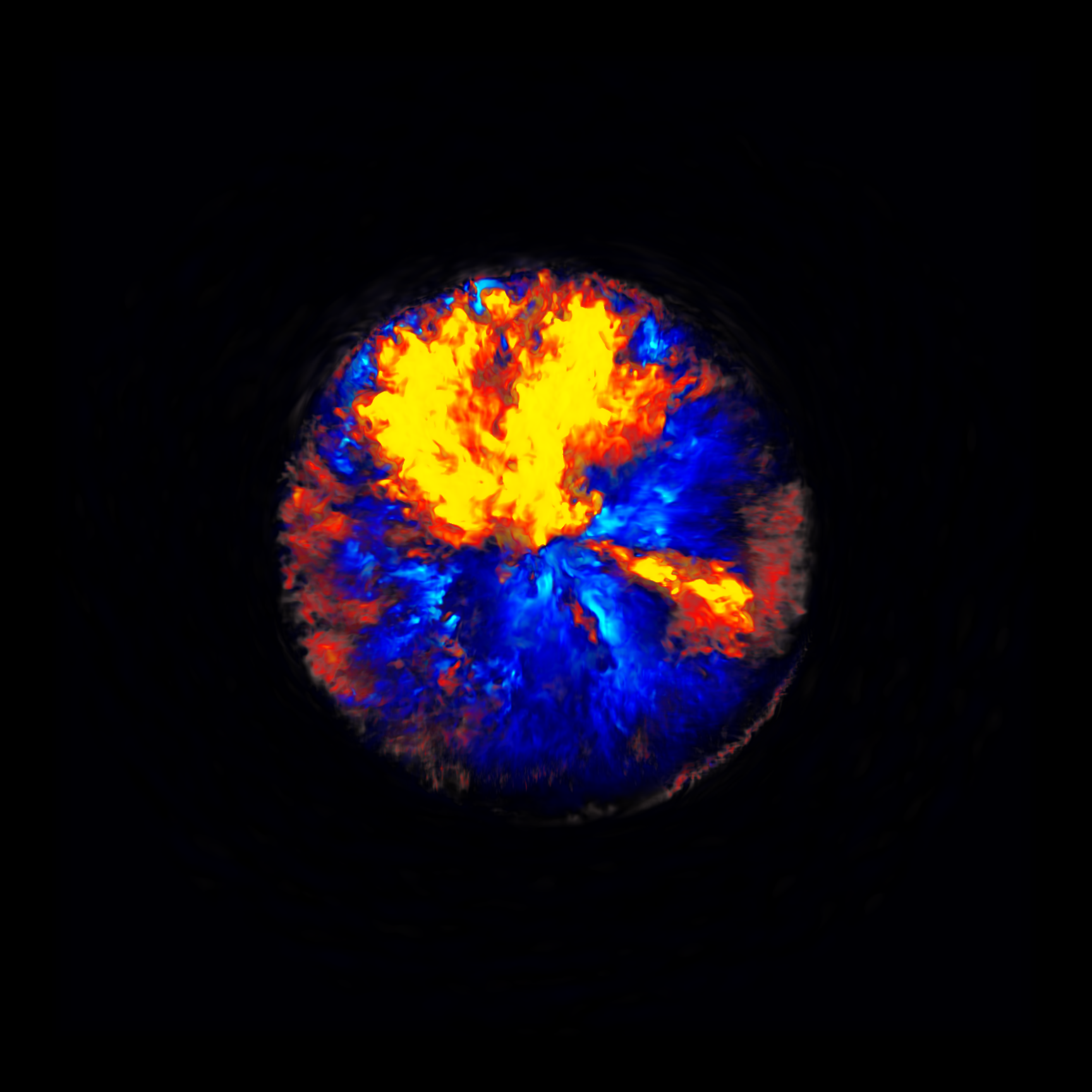}};
        \begin{scope}[x={(image.south east)},y={(image.north west)}]
            \draw[white,-Bar,-Triangle, line width=0.4mm] (0.5,0.9) -- (0.8235,0.9) node[midway,above] {22 Mm};
            \draw[white,dashed] (0.824,0.9) -- (0.824,0.5);
            \draw[white,-Bar,-Triangle, line width=0.4mm] (0.5,0.1) -- (0.882,0.1) node[midway,below] {26 Mm};
            \draw[white,dashed] (0.882,0.1) -- (0.882,0.5);
        \end{scope}
    \end{tikzpicture}
    \caption{Centre-plane slice rendering of run W20 (CPM setup, 1728$^3$ grid) at dump~1300 ($t=31.4\,$h). {\it Left}: magnitude of the tangential velocity component $|U_t|$ (i.e., perpendicular to the radial direction), with dark blue, turquoise, yellow, red, and dark red representing a sequence of increasing velocities. {\it Right}: radial velocity $U_r$, with blue colours representing inward-moving flows and red-orange colours outward-moving flows. These renderings were generated to qualitatively visualize the important features of our simulations. The full simulation domain is not shown here: only a $68\,{\rm Mm} \times 68\,{\rm Mm}$ region is displayed. White arrows indicate physical distances from the centre of the star. High-resolution movies are available at \url{https://www.ppmstar.org}.}
    \label{fig:W20bobs}
\end{figure*}

\begin{figure*}
    \centering
    \begin{tikzpicture}
        \node[anchor=south west,inner sep=0] (image) at (0,0) {\includegraphics[trim={7cm 7cm 7cm 7cm},clip,width=0.49\linewidth]{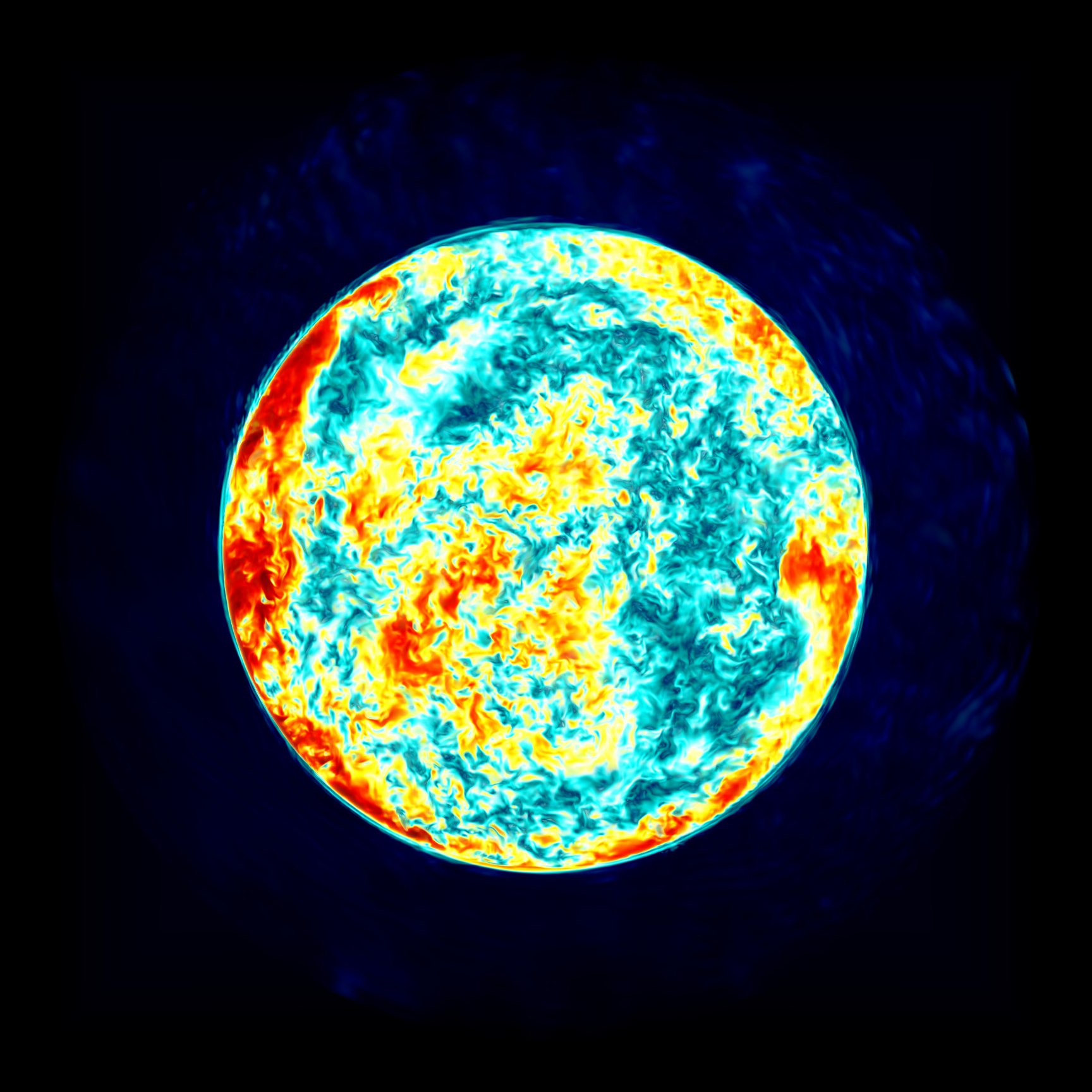}};
        \begin{scope}[x={(image.south east)},y={(image.north west)}]
            \draw[white,-Bar,-Triangle, line width=0.4mm] (0.5,0.9) -- (0.882,0.9) node[midway,above] {26 Mm};
            \draw[white,dashed] (0.882,0.9) -- (0.882,0.5);
        \end{scope}
    \end{tikzpicture}
    \begin{tikzpicture}
        \node[anchor=south west,inner sep=0] (image) at (0,0) {\includegraphics[trim={7cm 7cm 7cm 7cm},clip,width=0.49\linewidth]{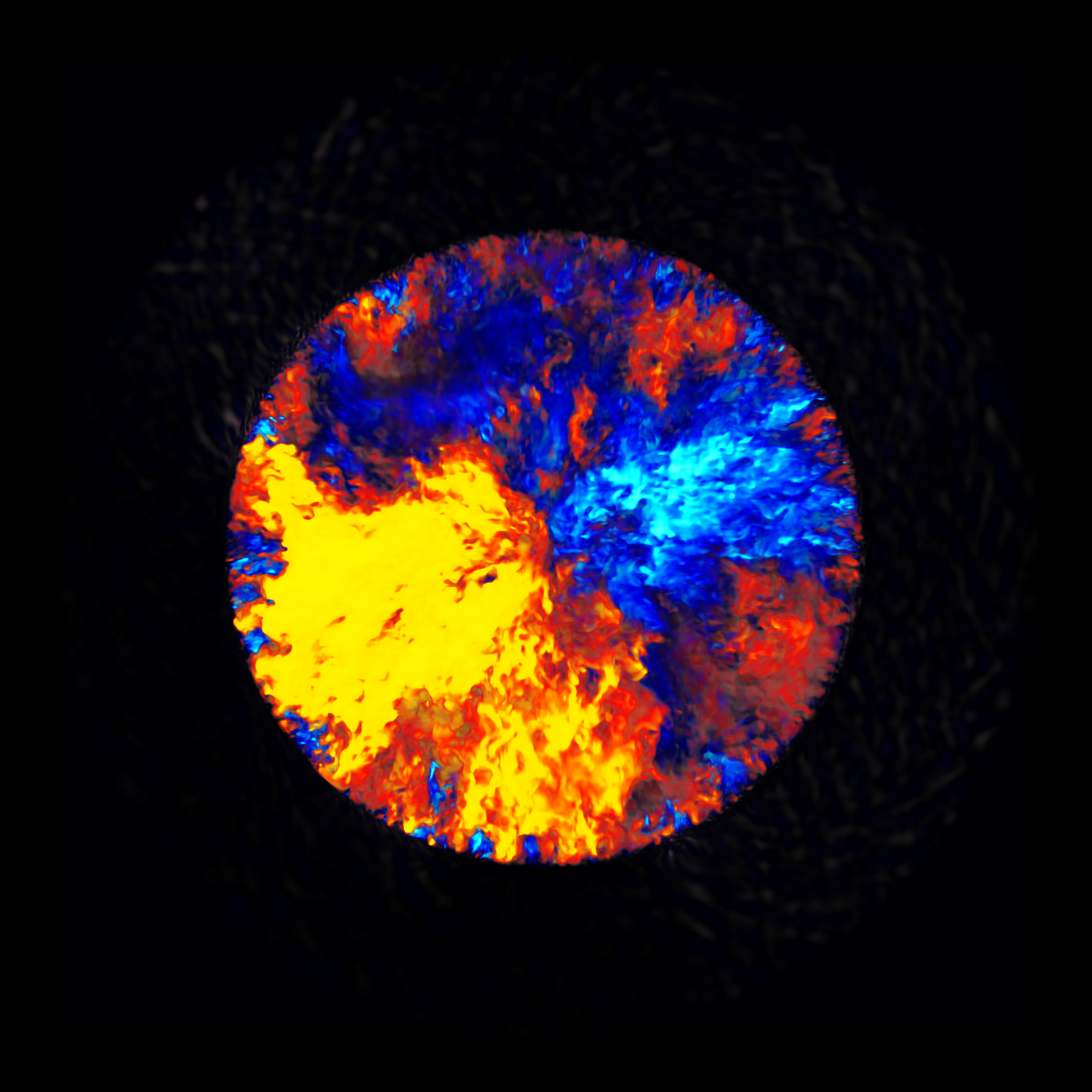}};
        \begin{scope}[x={(image.south east)},y={(image.north west)}]
            \draw[white,-Bar,-Triangle, line width=0.4mm] (0.5,0.9) -- (0.882,0.9) node[midway,above] {26 Mm};
            \draw[white,dashed] (0.882,0.9) -- (0.882,0.5);
        \end{scope}
    \end{tikzpicture}
    \caption{Identical to Figure~\ref{fig:W20bobs} but for run W21 (PM setup, 1728$^3$ grid) at dump~1300 ($t=31.4\,$h). The image scale is identical to that of Figure~\ref{fig:W20bobs}.}
    \label{fig:W21bobs}
\end{figure*}

We first investigate the qualitative behaviour of the flow in our simulations. Figures~\ref{fig:W20bobs} and~\ref{fig:W21bobs} are center-plane slice renderings of the tangential\footnote{By tangential we mean in the plane perpendicular to the radial direction.} (left panels) and radial (right panels) velocity components in our high-resolution $1728^3$ CPM and PM simulations. The most striking feature of these renderings is the large dipole-like structure that dominates the convective core, a finding that echoes previous 3D simulations of core convection \citep[e.g.,][]{gilet2013,anders2023,blouin2023b,herwig2023}. In Figure~\ref{fig:W20bobs}, we can clearly see in bright orange the large plume travelling from the centre and toward the north in the radial velocity rendering. When it hits the convective boundary, this rising plume is split into two diverging flows travelling in the tangential direction, which forms the characteristic horseshoe-like structure in the horizontal velocity rendering. As described in \cite{herwig2023}, these two tangential flows ultimately separate from the boundary due to their opposing pressure gradients when they eventually travel towards one another, thereby forming an inward moving plume. The same general behaviour can be observed in Figure~\ref{fig:W21bobs} for the PM setup. It takes a few convective turnover timescales for this dipolar circulation pattern to establish itself, but once it does, it remains a persistent characteristic of the flow. Note however that the dipole structure is not fixed: the movies available at \url{https://www.ppmstar.org} clearly show that its intensity and orientation fluctuate with time.

A comparison of the tangential velocity renderings of Figures~\ref{fig:W20bobs}-\ref{fig:W21bobs} immediately reveals the very different nature of the convective boundary region in both setups. In the PM case (Figure~\ref{fig:W21bobs}), there is a sharp transition between the high convective velocities of the core and the much smaller velocities that characterize the stable envelope (which are barely visible in these renderings). In contrast, in the CPM simulations (Figure~\ref{fig:W20bobs}), the semiconvection zone imprints a region of moderate velocities between the high velocities of the convective core and low velocities of the stable envelope. It is evident from the ring-like structure of the flow in this region that the semiconvection zone is not dominated by turbulent convective motions. This will be investigated in more detail in Section~\ref{sec:nature}.

\subsection{Convergence with respect to the grid resolution}
\label{sec:resolution}
\begin{figure*}
    \centering
    \includegraphics[width=0.49\linewidth]{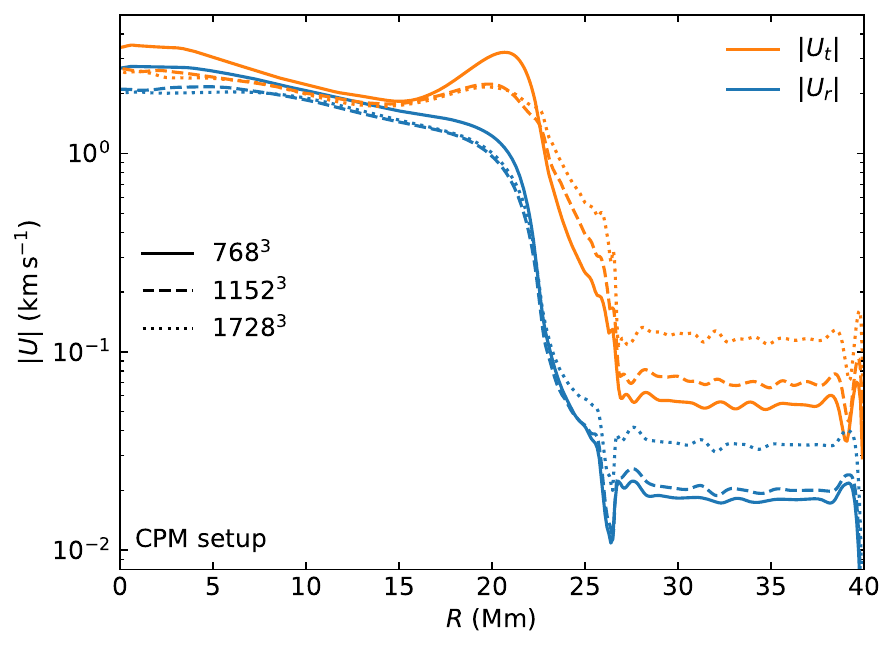}
    \includegraphics[width=0.49\linewidth]{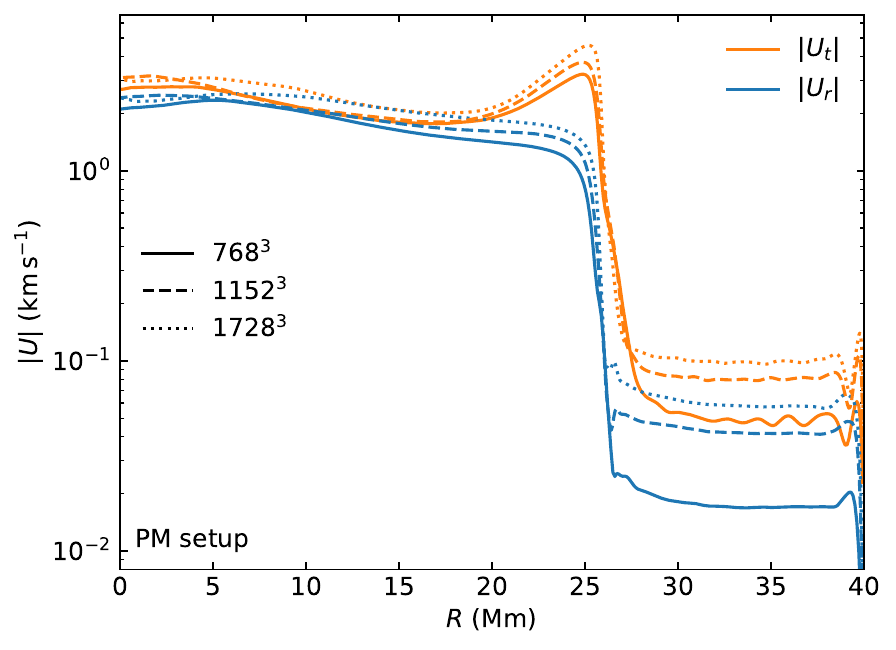}
    \caption{Spherically averaged tangential (orange) and radial (blue) velocity profiles for the CPM (left) and PM (right) setups for different grid resolutions. Simulations W12, W13, W20, W21, W22 and W23 were used to generate this figure ($\log L/L_{\star}=5$ heating). The velocity profiles were obtained by averaging over the last 400 dumps of each simulation.}
    \label{fig:velcomp}
\end{figure*}

In Figure~\ref{fig:velcomp} we compare the spherically averaged velocity profiles of our $\log L/L_{\star}=5$ simulations performed using three different grid resolutions. For both setups, our results indicate that the properties of the flow in the convective core and in the boundary region are already well converged with respect to the grid resolution at $1152^3$ since there is little difference between the $1152^3$ and $1728^3$ cases. The situation is less favourable in the envelope. With the CPM setup, there is no sign of convergence of the envelope velocities with respect to the simulation grid. The situation is better in the case of the PM setup, where the decreasing velocity difference between successive grid refinements suggests that the velocities are approaching convergence.

The slower convergence of the velocities in the envelope compared to the core (or lack of convergence in the CPM case) is at least partially due to the smaller velocities in that region of the star. At the $\log L/L_{\star}=5$ heating rate shown in Figure~\ref{fig:velcomp}, the Mach number in the envelope is only of the order of $10^{-4}$, a challenging regime for an explicit gas dynamics code. This also implies that a better convergence under grid refinement is expected for our simulations with higher heating luminosities, as the flow in the envelope will be more vigorous (Section~\ref{sec:heating}). In any case, this issue is not a major concern in what follows, as the central goal of this work is to study the boundary region where for both setups the velocities converge much better under grid refinement.

We have seen that our PM simulations develop a fully convective core separated from the stable He envelope by a well-defined boundary (Figures~\ref{fig:W21bobs}--\ref{fig:velcomp}). It is unsurprising to find a large fully mixed core with the PM prescription, but it is still interesting to note that we cannot identify any impact on the gas dynamics resulting from the presence of a region in the core where the Schwarzschild criterion is barely satisfied. Indeed, at $R = 21.3\,{\rm Mm}$, $\nabla_{\rm rad} - \nabla_{\rm ad}$ reaches its minimum in the convective core of just 0.002, and yet there is no obvious slowdown of the convective motions in that region (right panel of Figure~\ref{fig:velcomp}). The global dipolar circulation pattern is completely oblivious to the presence of this minimum. At least at the heating rates of our simulations, this result confirms that it is consistent to assume full mixing in regions that have $\nabla_{\rm rad} - \nabla_{\rm ad} > 0$ by an arbitrarily small margin, as assumed in the maximum overshoot prescription and with \code{MESA}'s PM scheme.

\subsection{Heating series}
\label{sec:heating}

\begin{figure*}
    \centering
    \includegraphics[width=0.49\linewidth]{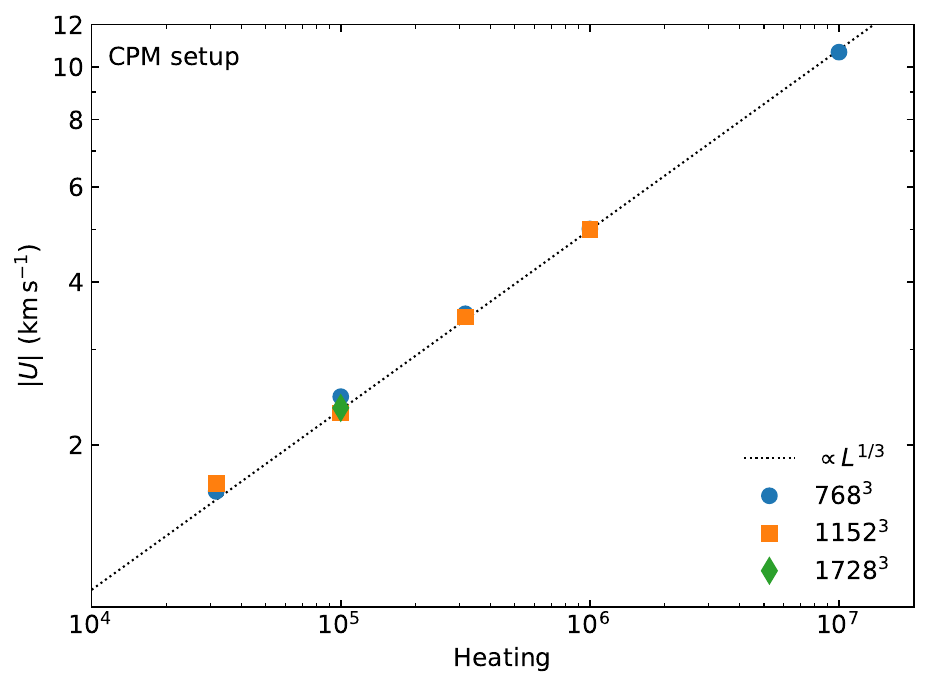}
    \includegraphics[width=0.49\linewidth]{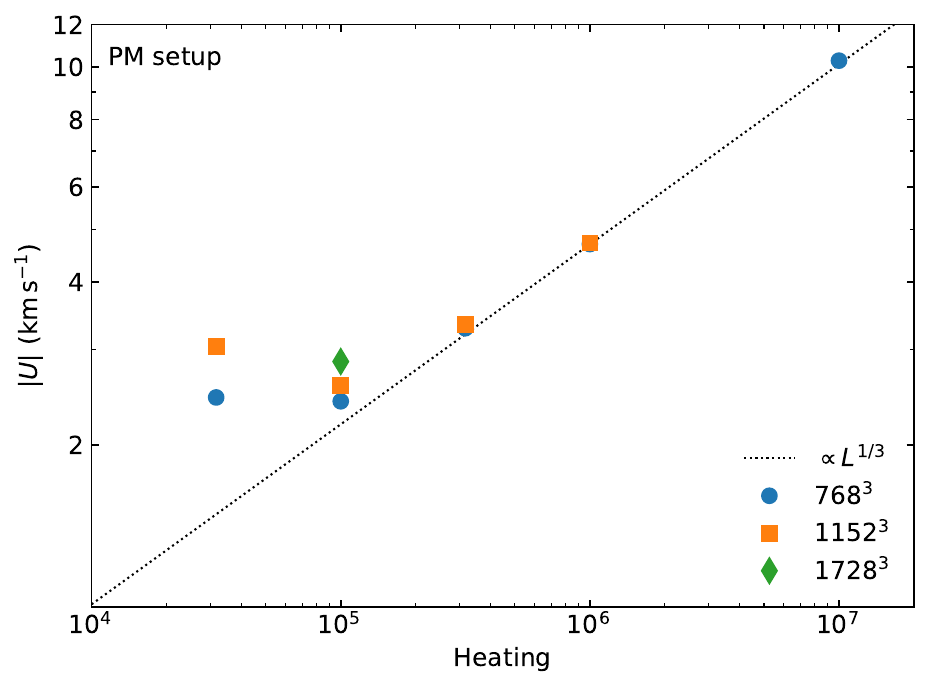}
    \caption{Average velocity in the convective core at $R=15\,$Mm as a function of the heating factor $L/L_{\star}$. Results for the CPM setup are shown in the left panel and those for the PM setup are shown in the right panel. The different symbols indicate the grid resolution. For reference, the dotted line shows a $L^{1/3}$ power law. All simulations listed in Table~\ref{tab:runs} were used to generate this figure. The rms velocities were averaged over the last 15\,h of each simulation.}
    \label{fig:heating_cvz}
\end{figure*}

\begin{figure*}
    \centering
    \includegraphics[width=0.49\linewidth]{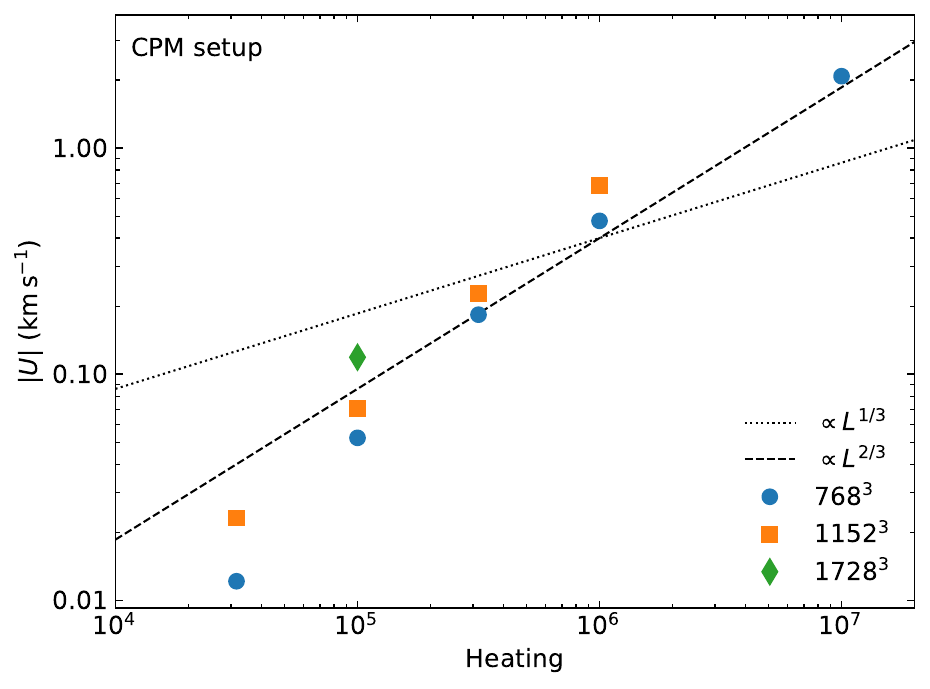}
    \includegraphics[width=0.49\linewidth]{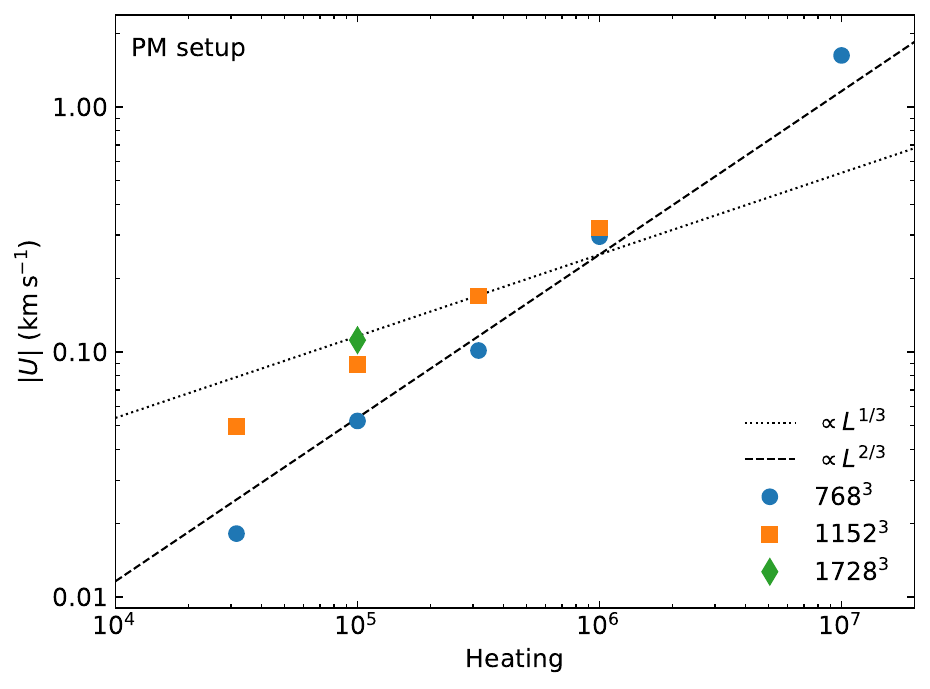}
    \caption{Same as Figure~\ref{fig:heating_cvz}, but this time looking at the velocities at $R=33\,$Mm in the stable envelope.}
    \label{fig:heating_igw}
\end{figure*}

As explained in Section~\ref{sec:ppmstar}, all our simulations are performed using heating luminosities that far exceed the nominal luminosity of the star. It is therefore important to understand how the properties of the flow scale with respect to the heating rate in order to properly extrapolate to the nominal case. To do so, we show in Figure~\ref{fig:heating_cvz} the rms velocity $|U|$ in the convective core at $R=15\,$Mm as a function of $L/L_{\star}$. For the CPM setup, we precisely recover a $L^{1/3}$ scaling law, as expected based on previous results of 3D hydrodynamics simulations of convection in stars \citep[e.g.,][]{porter2000,muller2016,jones2017,baraffe2021,herwig2023}. Note also the excellent convergence with respect to the grid resolution: there is virtually no difference between results obtained using a 768$^3$, 1152$^3$ or 1728$^3$ grid. However, the situation is different for the PM setup. First, departures from the $L^{1/3}$ power law appear at $\log L/L_{\star} \leq 5$. Second, while there is excellent convergence with respect to the grid resolution at $\log L/L_{\star} > 5$ and good convergence at $\log L/L_{\star} = 5$ (consistent with Figure~\ref{fig:velcomp}), there is a more significant difference between the 768$^3$ and 1152$^3$ runs at $\log L/L_{\star} = 4.5$. Given that the convective velocities are similar for both setups, it is surprising to see this departure from the expected scaling law in the PM setup while the expected behaviour is recovered down to at least $\log L/L_{\star} = 4.5$ in the CPM case. Nevertheless, the fact that the $L^{1/3}$ power law still applies for $\log L/L_{\star} \geq 5$ supports the use of this scaling relation to extrapolate to nominal luminosity using the $\log L/L_{\star} \geq 5$ simulations.

Figure~\ref{fig:heating_igw} repeats the same exercise in the stable envelope at $R=33\,$Mm. Based on the results of \cite{herwig2023}, we expect a $L^{2/3}$ scaling relation in this region where internal gravity wave (IGW) motions excited by the convective core characterize the flow (we will demonstrate the IGW-dominated nature of the flow in Section~\ref{sec:nature}). This is indeed what we recover for both setups at high luminosities, and as expected we observe a departure from this scaling relation at lower luminosities. In addition, there is a systematic offset between the $768^3$ and $1152^3$ scaling laws. This resolution dependence is a reflection of the fact that the velocities in the envelope are not converged with respect to the grid resolution (Section~\ref{sec:resolution}). 

\subsection{Power spectra}
\label{sec:nature}
\begin{figure*}
    \centering
    \includegraphics[width=0.495\linewidth]{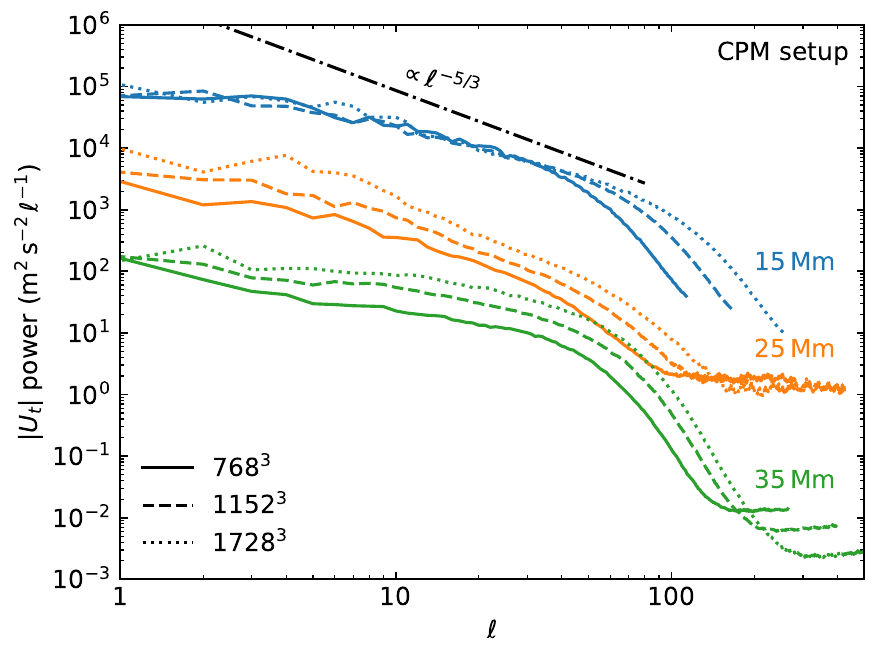}
    \includegraphics[width=0.495\linewidth]{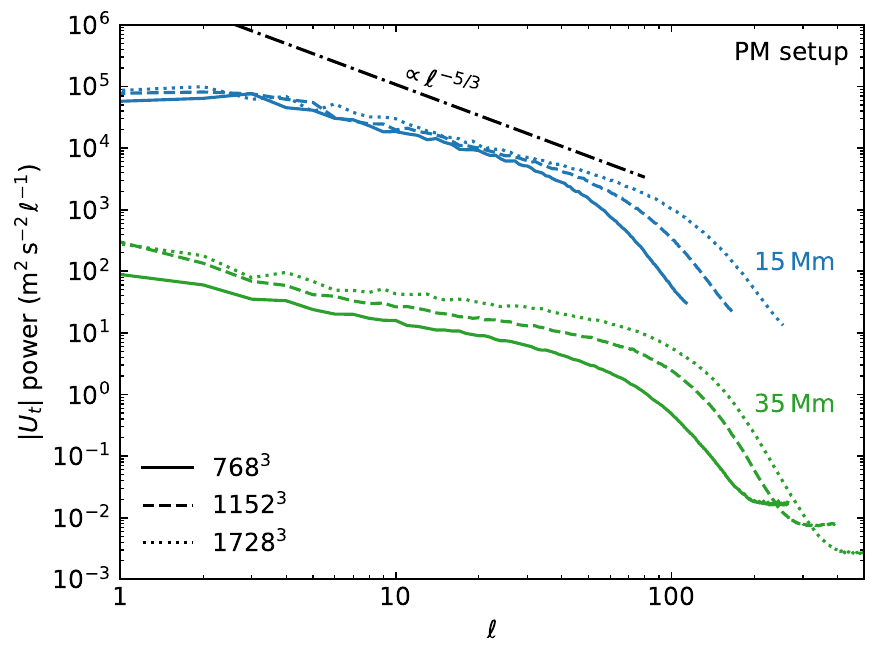}
    \includegraphics[width=0.495\linewidth]{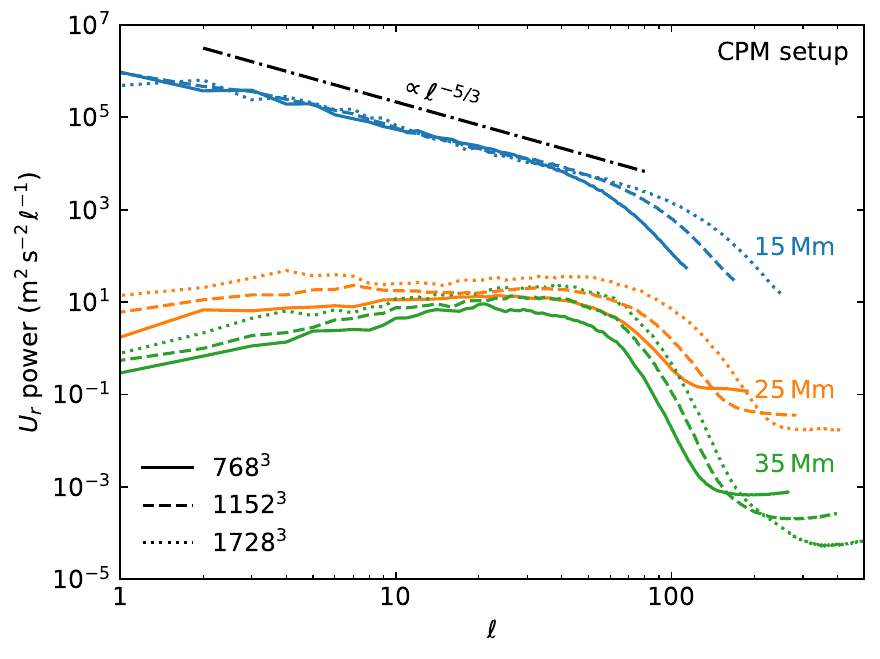}
    \includegraphics[width=0.495\linewidth]{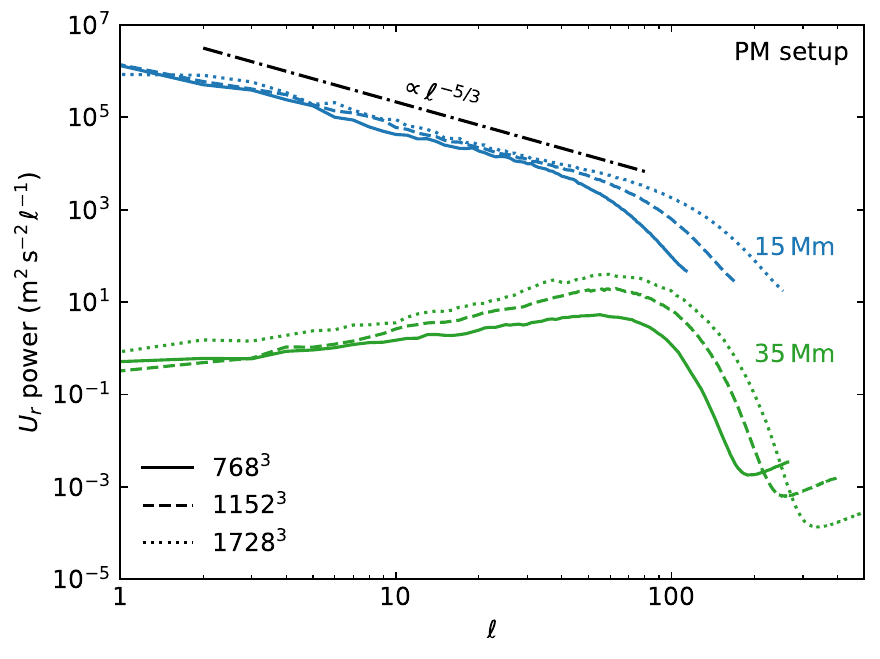}
    \caption{Power spectra of the tangential velocity magnitude $|U_t|$ (top row) and radial velocity $U_r$ (bottom row) taken at different radii in the simulations and for different grid resolutions. The spectra in blue are in the convective core, those in green are in the envelope, and for the CPM setup (left column) those in orange are in the semiconvective region. The spectra were calculated by averaging over dumps 800 to 1000 of our $\log L/L_{\star}=5$ simulations (runs W12, W13, W20, W21, W22 and W23).}
    \label{fig:power}
\end{figure*}

We now go beyond spherical averages of the velocity field and investigate how much power the flow contains across different length scales. Figure~\ref{fig:power} shows the power spectra of the tangential velocity magnitude $|U_t|$ ($S[|U_t|]$, top row) and radial velocity $U_r$ ($S[U_r]$, bottom row), where $S[]$ is the power spectrum operator. Power spectra are displayed for different radii in the star and for different Cartesian grid resolutions. The power is binned as a function of the angular degree $\ell$ and is calculated using \code{SHTools} \citep{wieczorek2018}. The maximum $\ell$ value shown in Figure~\ref{fig:power} depends on the radius at which the power spectrum is calculated. A larger radius or grid resolution allows us to capture smaller scale features, since the angular resolution of the projection of the Cartesian grid on a sphere increases. Note that the spectra are calculated using the filtered briquette data, which is down-sampled by a factor of 4 in each direction.

Both for $|U_t|$ and $U_r$, the power spectra in the convective core (blue lines, $R=15\,$Mm in Figure~\ref{fig:power}) is close to the expected Kolmogorov $\ell^{-5/3}$ cascade from large-scale modes ($\ell=1-2$) down to very small scales ($\ell \sim 100$ for high-resolution runs) where dissipation takes place. The convective power spectra are consistently extended to higher $\ell$ values when the grid resolution is increased. This reflects the fact that the turbulent cascade extends down to the smallest spatial scales resolved in the simulation. In the stable envelope (green lines, $R=35\,$Mm) and in the semiconvective region (orange lines, $R=25\,$Mm), the $U_r$ power spectra display a very different shape compared to the convective core, with a shallow increasing slope up to $\ell \sim 60$ followed by a rapid drop at higher $\ell$. This shape is reminiscent of that found in the stable layers of our recent main-sequence and red giant branch \code{PPMstar} simulations \citep{herwig2023,blouin2023a} and hints at the IGW-dominated nature of the flow at those radii. Contrarily to the convective power spectra, the power at all $\ell \lesssim 200$ continuously increases upon grid refinement. This behaviour is consistent with the previously noted absence of convergence for the velocities in the stable envelope of the CPM runs (Section~\ref{sec:resolution}). We speculate that this increase in IGW power could be the result of the extension of the convective power spectra to larger $\ell$ values with increasing resolution. IGWs are possibly mainly excited by small-scale convective motions close to the core boundary, which would imply stronger IGW motions at high resolutions due to the enhanced power at high $\ell$ in the convective spectra.

For both setups, the power spectra at $R=35\,$Mm are approaching convergence for $\ell \gtrsim 100$: the separation between the $1152^3$ and $1728^3$ spectra is smaller than that between the $768^3$ and $1152^3$ spectra. This is to be contrasted with the behaviour observed in the convective layers, where the power spectrum always extends to larger $\ell$ values upon grid refinement. This is most likely the result of radiative damping. Wave velocities are expected to be damped when radiative diffusion is taken into account, as the temperature of oscillating fluid parcels is equilibrated with their surroundings. This damping is stronger at large $\ell$ \citep{zahn1997}, since small fluid parcels can lose their heat more quickly than large ones. This is consistent with the saturation of the IGW power spectrum observed in Figure~\ref{fig:power}. This result implies that radiative damping in our simulations only affects very small spatial scales ($\ell \gtrsim 100$) that require large Cartesian grids to be properly resolved (above $1152^3$).

\begin{figure*}
    \centering
	\includegraphics[width=\linewidth]{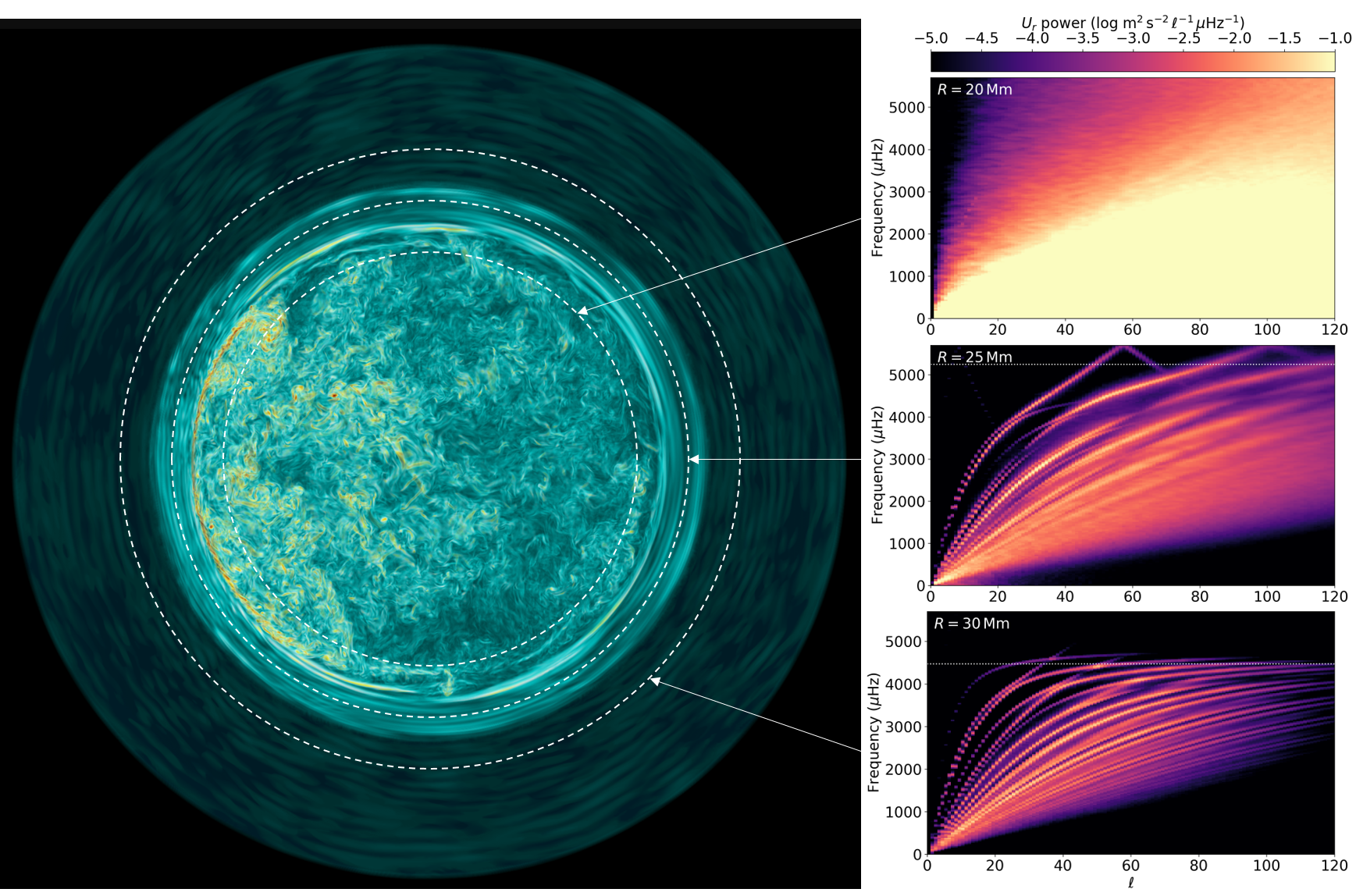}
    \caption{{\it Left}: Centre-plane slice rendering of the vorticity magnitude $|\nabla \times U|$ of run W20 (CPM setup, 1728$^3$ grid) at dump~840 ($t=20.3\,$h). Dark blue, turquoise, yellow, red, and dark red represent a sequence of increasing vorticity magnitudes. Unlike in Figure~\ref{fig:W20bobs} and~\ref{fig:W21bobs}, the full simulation domain is shown. The white circles indicate the radii at which power spectra were calculated. {\it Right}: Power spectra of the radial velocity components at $R=20,25,30\,$Mm. The spectra were calculated over dumps 480 to 880 ($t=11.6-21.3$h). The local value of the Brunt--V\"ais\"al\"a frequency is shown by the dotted horizontal line.}
    \label{fig:scvz-kw}
\end{figure*}

\begin{figure}
\centering
	\includegraphics[width=\columnwidth]{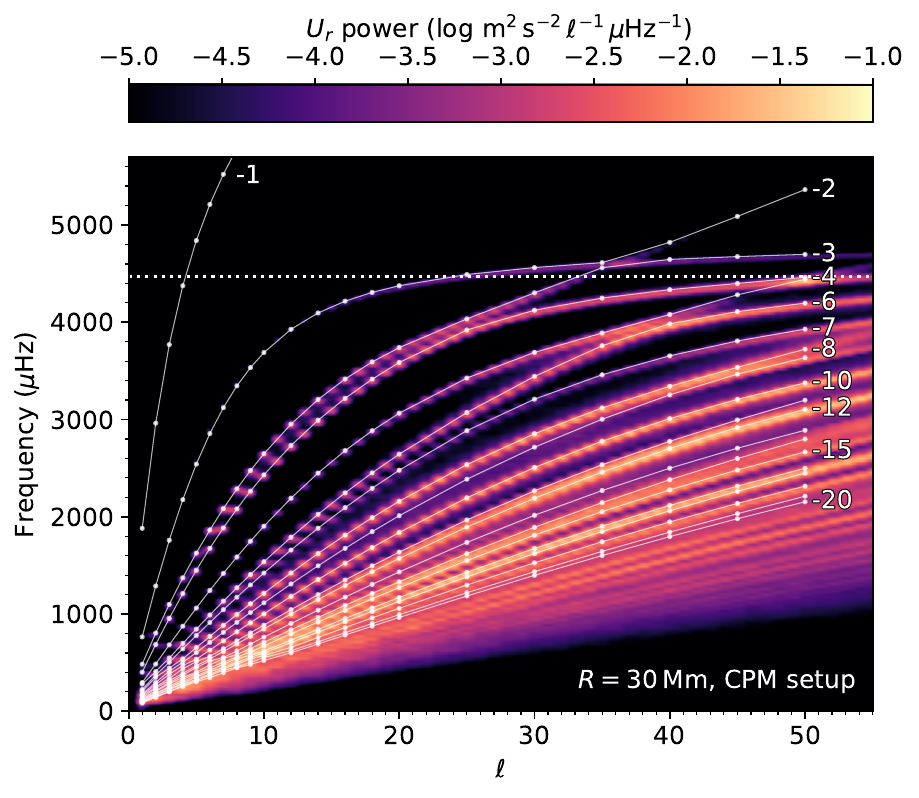}
	\caption{Power spectra of the radial velocity component at $R=30\,$Mm for run W20 calculated over dumps 480 to 880 ($t=11.6-21.3$h). The local value of the Brunt--V\"ais\"al\"a frequency is shown by the dotted horizontal line. The overlaid solid white lines show the eigenmodes predicted by \code{GYRE} for the stellar structure corresponding to that of run W20. The \code{GYRE} analysis was carried out using the spherically averaged structures of dumps 480 and 880: the frequencies shown here are averages of these two calculations. The labels next to the white lines indicate the radial orders $n$ of the modes (their negative signs indicate that they are $g$ modes).}
	\label{fig:kw-gyre}
\end{figure}

We have so far been discussing various properties of the IGWs in the stable envelope, but without explicitly demonstrating that IGWs indeed dominate fluid motions in that region of the star. We remediate this issue in Figure~\ref{fig:scvz-kw}. We show $U_r$ power spectra taken at three different radii in run W20 (CPM setup, 1728$^3$ grid). Here, the power is binned both as a function of $\ell$ and the temporal frequency (the methodology used to calculate these spectra is detailed in \citealt{thompson2022}). The power spectra calculated in the convective core shows power spread out over a wide range of spatial and temporal frequencies with no clear structure, as expected for turbulent motions. In contrast, the power spectra calculated in the stable envelope (bottom panel) displays distinct ridges constituted of discrete modes. These discrete modes quickly disappear for frequencies that surpass the local Brunt--V\"ais\"al\"a frequency (represented here by a thin dotted line). This is precisely what is expected for IGWs, which are damped when the Brunt--V\"ais\"al\"a frequency is exceeded. Figure~\ref{fig:kw-gyre} further demonstrates the IGW nature of these modes. Here, we zoom in on the low-$\ell$ portion of the bottom panel of Figure~\ref{fig:scvz-kw} and overlay eigenmodes predictions from the stellar oscillation code \code{GYRE} \citep{townsend2013,townsend2018}. This analysis is performed as in \cite{thompson2022} and is based on spherically averaged radial profiles from run W20. Two \code{GYRE} calculations were performed: one using the structure corresponding to the beginning of the series of dumps used in calculating the power spectrum from the simulation data and one using the last dump of that series. The frequencies shown in Figure~\ref{fig:kw-gyre} are the averages of these two calculations. The excellent agreement between the \code{GYRE} predictions definitely confirms that these modes are IGWs, and it even allows the identification of individual radial orders.

Going back to Figure~\ref{fig:scvz-kw}, we now focus on the particularly interesting power spectrum of the semiconvection zone (middle panel). We see the same discrete ridges as in the stable envelope, signalling the presence of IGWs and indicating that the semiconvective layers are dominated by IGW motions. The power distribution is admittedly more smeared out than in the stable envelope where the ridges are well separated, but this is presumably due to shorter mode lifetimes. The lack of power at high $\ell$ and low frequencies is further indication of the absence of convective motions in this region. We will investigate the potential implications of the IGW-dominated nature of the semiconvection zone on CBM in Section~\ref{sec:IGWmixing}.

\section{On the nature of the convective boundary}
\label{sec:boundary}
Having described the main properties of the flow, established the numerical convergence properties of our simulations and obtained scaling relations, we now shift our focus to the CHeB CBM problem introduced in Section~\ref{sec:intro}. As a reminder, we have performed two series of simulations using two different initial setups with different CBM prescriptions because the accurate boundary mixing scheme is unknown. Can we use these simulations to infer the best CBM prescription to employ? In Section~\ref{sec:evol}, we explore the most direct approach to tackle this question, namely investigating the evolution of the stratification in the boundary region.

\subsection{Long-term evolution of the convective boundary}
\label{sec:evol}
\begin{figure*}
    \centering
	\includegraphics[width=0.495\linewidth]{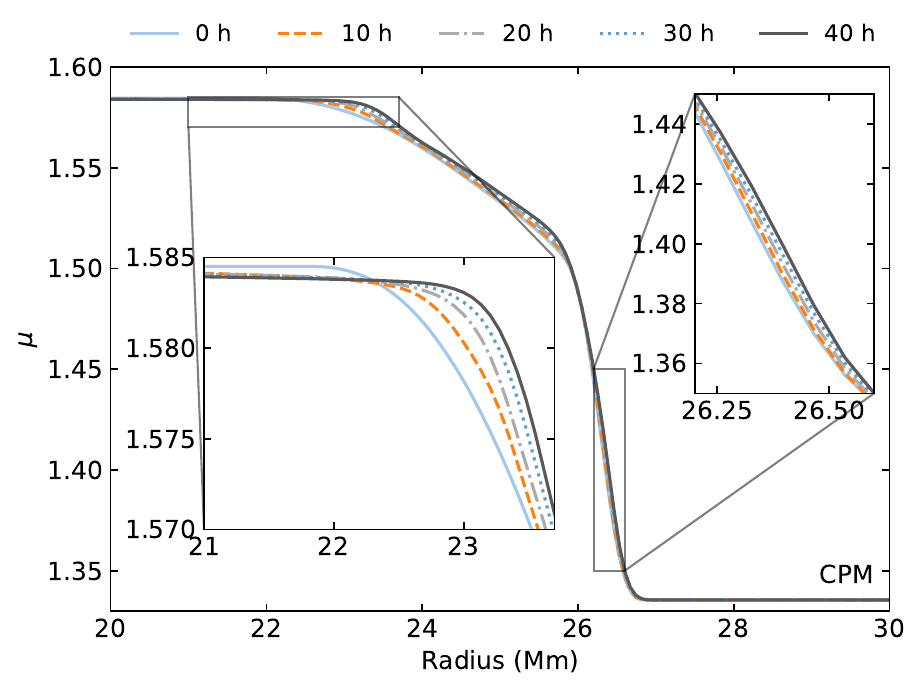}
    \includegraphics[width=0.495\linewidth]{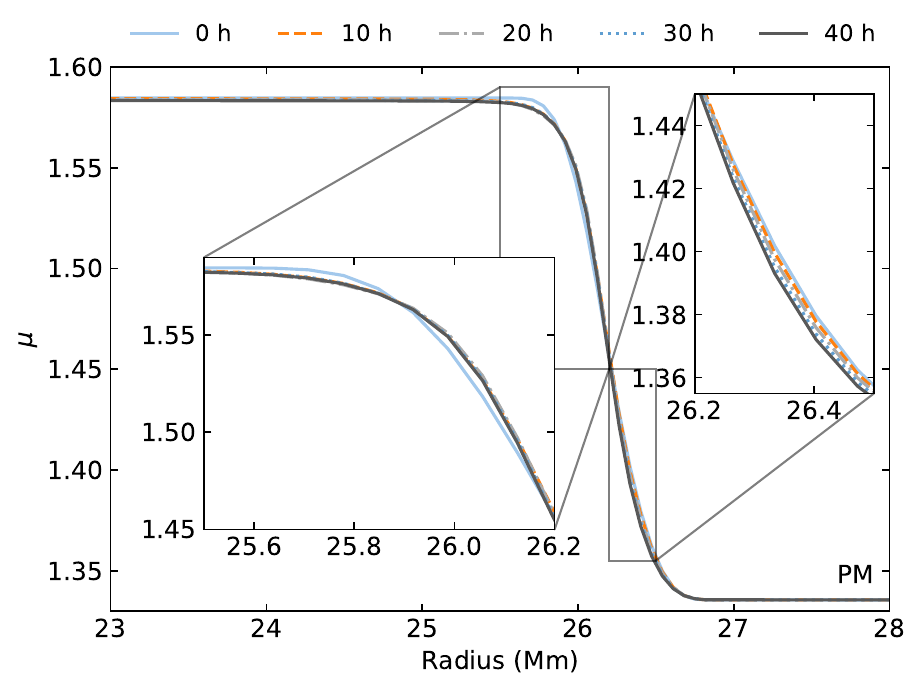}
    \caption{Evolution of the mean molecular weight profile in the boundary region for a CPM (left panel, run W28) and a PM (right panel, run W29) simulation. These runs employed a 1152$^3$ grid and a  $\log L/L_{\star}=6$ heating rate. The legends above the panels indicate the simulation times at which the profiles are shown.}
    \label{fig:evol}
\end{figure*}

Figure~\ref{fig:evol} shows the time evolution of the $\mu$ profile in the boundary region for our $\log L/L_{\star}=6$ simulations performed on a 1152$^3$ grid. For the CPM setup, we see that both the convective--semiconvective (left inset in the left panel) and semiconvective--radiative (right inset in the left panel) boundaries appear to be slowly migrating outward with time. The boundary is also changing for the PM setup, where the $\mu$ profile is steepening with the outer portion of the boundary moving inward (right inset in the right panel). These changes are admittedly small compared to the 0.07\,Mm grid cell size of these simulations, but they nevertheless constitute a robust result that we recover at other grid resolutions and heating rates. 

Note that these migrations do not show sign of slowing down and approaching equilibrium after 40\,h of simulation time. At the boundary, the radiative diffusivity is $D_{\rm rad} \sim 10^{11}\,{\rm cm}^2\,{\rm s}^{-1}$ in our $\log L/L_{\star}=6$ simulations, which implies that heat diffusion had time to operate over a length scale of only $\sqrt{D_{\rm rad} \cdot 40\,{\rm h}} \sim 1\,{\rm Mm}$. Hence, there is not enough time to reinstate thermal equilibrium following a local perturbation in the thermal structure of the stable layers due to a displacement of the convective boundary. It is therefore unsurprising that the boundary is still evolving at the end of our simulations, and reaching a stable boundary would demand much longer simulations \citep{anders2022,herwig2023,mao2023}. Nevertheless, it is tempting to extrapolate these trends to find the final equilibrated state towards which the simulations are evolving. For example, the apparent outward migration of the CPM setup may imply a larger mixed core than the one present in the initial \code{MESA} stratification.

However, such extrapolations are hazardous as it is not clear that these trends also apply to the real star. We have seen in Section~\ref{sec:mapping} that the mapping between the \code{MESA} stratifications and the \code{PPMstar} base states is not perfect. The main problem is the omission of electron degeneracy pressure and ion--ion nonideal interactions in the equation of state currently employed by \code{PPMstar}, which leads to a 10\% offset of the pressure profile. This inaccuracy affects the whole stratification, including the temperature and opacity profiles. This in turn perturbs the thermal balance of the star in a way that can be expected to shift its convective boundary. Another way to see this is to realize that a hypothetical \code{MESA} calculation with a different equation of state and/or opacity tables would yield a different convective boundary, since the Schwarzschild criterion would not be satisfied at the same radius. Hence, the small convective boundary reconfigurations observed in Figure~\ref{fig:evol} can at least partially be attributed to the response of the simulations to a change in the equation of state that has induced a state of thermal imbalance in the star. There is no obvious way to separate this behaviour from the (more interesting) 3D hydrodynamical response to the initial \code{MESA} base state, which is thermally balanced within the MLT framework.

All things considered, it is not currently possible to directly extrapolate the long-term evolution of our simulations to infer the ``correct'' stratification at the convective boundary of CHeB stars. Nevertheless, we will see in Sections~\ref{sec:IGWmixing} and~\ref{sec:plumes} that we can gain further insights on the evolution of the convective boundary in our CPM simulations by isolating individual mixing processes through measurements of the diffusivity profile.

\subsection{IGW mixing}
\label{sec:IGWmixing}
We have described in Section~\ref{sec:nature} how the semiconvection zone of our CPM simulations is dominated by strong IGW motions. Our recent simulations of core convection in massive main-sequence stars have revealed species mixing in stable layers with strong IGW motions, presumably because of IGW-induced mixing \citep{herwig2023}. A priori, the same phenomenon could conceivably occur inside the semiconvection zones of CHeB stars. If it is the case, then this would provide a mechanism to homogenize and destroy the semiconvective interface.

\begin{figure}
    \centering
	\includegraphics[width=\columnwidth]{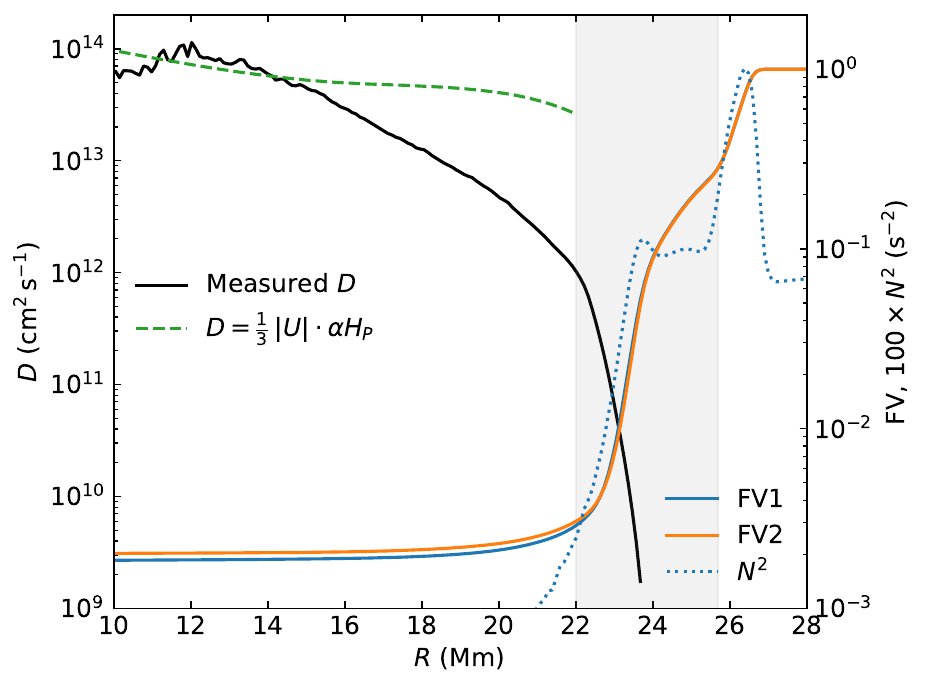}
    \caption{Measured diffusion coefficient (black solid line) in run W10 ($\log L/L_{\star}=6$), where $D$ was calculated by inverting the diffusion equation given the evolution of the FV profile between dumps 1300--1800 (blue solid line) and dumps 1800--2300 (orange solid line). Beyond 23.7\,Mm, diffusion coefficients become too small to be measured (the spherically averaged FV profiles are too similar). The Brunt--V\"ais\"al\"a is given by the blue dotted line, and the grey region is the semiconvection zone. For comparison, the MLT diffusion coefficient $D=\frac{1}{3} |U| \alpha H_P$ is shown in green, where we have assumed that $\alpha=0.25$ and that $|U|$ is given by the rms velocity profile in run W10.}
    \label{fig:diffusivity-IGW}
\end{figure}

We have measured the diffusion coefficient in our CPM simulations using the method described in \cite{jones2017} to invert the observed evolution of the FV profile, where FV is the fraction that represents the contribution of the envelope fluid to the two-fluid mixture.\footnote{The mean molecular weight of the two-fluid mixture is given by $\mu = {\rm FV} \cdot \mu_{\rm env} + (1- {\rm FV}) \cdot \mu_{\rm core}$.} The resulting diffusivity profile for a $\log L/L_{\star}=6$ run is shown in black in Figure~\ref{fig:diffusivity-IGW}. Deep inside the convective core, the measured diffusivity is consistent with a simple MLT prescription (green dashed line). $D$ becomes much smaller than the MLT value closer to the convective--semiconvective boundary, a well-known limitation of this simple prescription \citep{eggleton1972,jones2017,herwig2023,blouin2023b,blouin2023a}. Once we reach the semiconvection zone, $D$ drops precipitously and we cannot measure any mixing above $R=23.7\,{\rm Mm}$. The mixing measured near the convective--semiconvective boundary is due to the overshooting motions that we will explore further in Section~\ref{sec:plumes}. 

Beyond this boundary region, we measure no mixing in the semiconvection zone. This implies that despite their vigour, IGWs do not produce measurable mixing: we can establish an upper limit of $D_{\rm IGW} \sim 10^9\,{\rm cm}^2\,{\rm s}^{-1}$ in the semiconvection of our $\log L/L_{\star}=6$ simulations. Given the high heating rates of our simulations, we have every reason to believe that we are overestimating IGW mixing as we are overestimating IGW velocities by orders of magnitude. The upper limit on $D_{\rm IGW}$ implied by our simulations is therefore much smaller at nominal luminosity. In the absence of measurable IGW mixing in our simulations, it is of course impossible to establish a scaling relation for $D_{\rm IGW}$ that could be used to extrapolate to nominal luminosity. As an alternative, we can assume that the $D_{\rm IGW} \propto L^{4/3}$ relation found in the IGW-dominated convective boundary layers of the H-core burning simulations without radiation diffusion of \cite{herwig2023} still holds here. With this assumption, the upper limit on IGW mixing at nominal luminosity would be of only $\sim 10\,{\rm cm}^2\,{\rm s}^{-1}$. To evaluate whether this level of mixing could have an impact on the star, we estimate how long it would take to homogenize the semiconvective layers that are located above the radius where we can still measure $D$ in Figure~\ref{fig:diffusivity-IGW}. It would take more than $(2\,{\rm Mm})^2/D_{\rm IGW} \simeq 130\,{\rm Myr}$ to homogenize this 2\,Mm-thick shell. This is a timescale comparable to the entire CHeB lifetime: IGW mixing in the semiconvection zone is negligible assuming the $D_{\rm IGW} \propto L^{4/3}$ scaling relation is applicable to these simulations.

This result is consistent with the classical picture of semiconvection, where the diffusion coefficient in the semiconvective layers is taken to be \citep{langer1985}
\begin{equation}
    D_{\rm sc} = \alpha_{\rm sc} \frac{K}{6 c_{\scriptscriptstyle P} \rho} \frac{\nabla - \nabla_{\rm ad}}{\nabla_{\rm L}-\nabla},
\end{equation}
where $\alpha_{\rm sc}$ is a free parameter that controls the mixing timescale, $c_{\scriptscriptstyle P}$ is the heat capacity at constant pressure, and $\nabla_{\rm L}$ is the Ledoux gradient. Once an adiabatic stratification is reached ($\nabla-\nabla_{\rm ad}=0$, as is the case in the semiconvective region of our CPM setup), no further mixing takes place and the composition profile remains constant if secular changes to the structure are neglected. In this picture, the amplitudes of IGWs saturate before they become strong enough to overturn \citep[e.g.,][]{merryfield1995} and trigger nonlinear turbulent mixing.

\subsection{Convective overshooting}
\label{sec:plumes}
\begin{figure}
    \centering
     \begin{tikzpicture}
        \node[anchor=south west,inner sep=0] (image) at (0,0) {\includegraphics[width=\columnwidth]{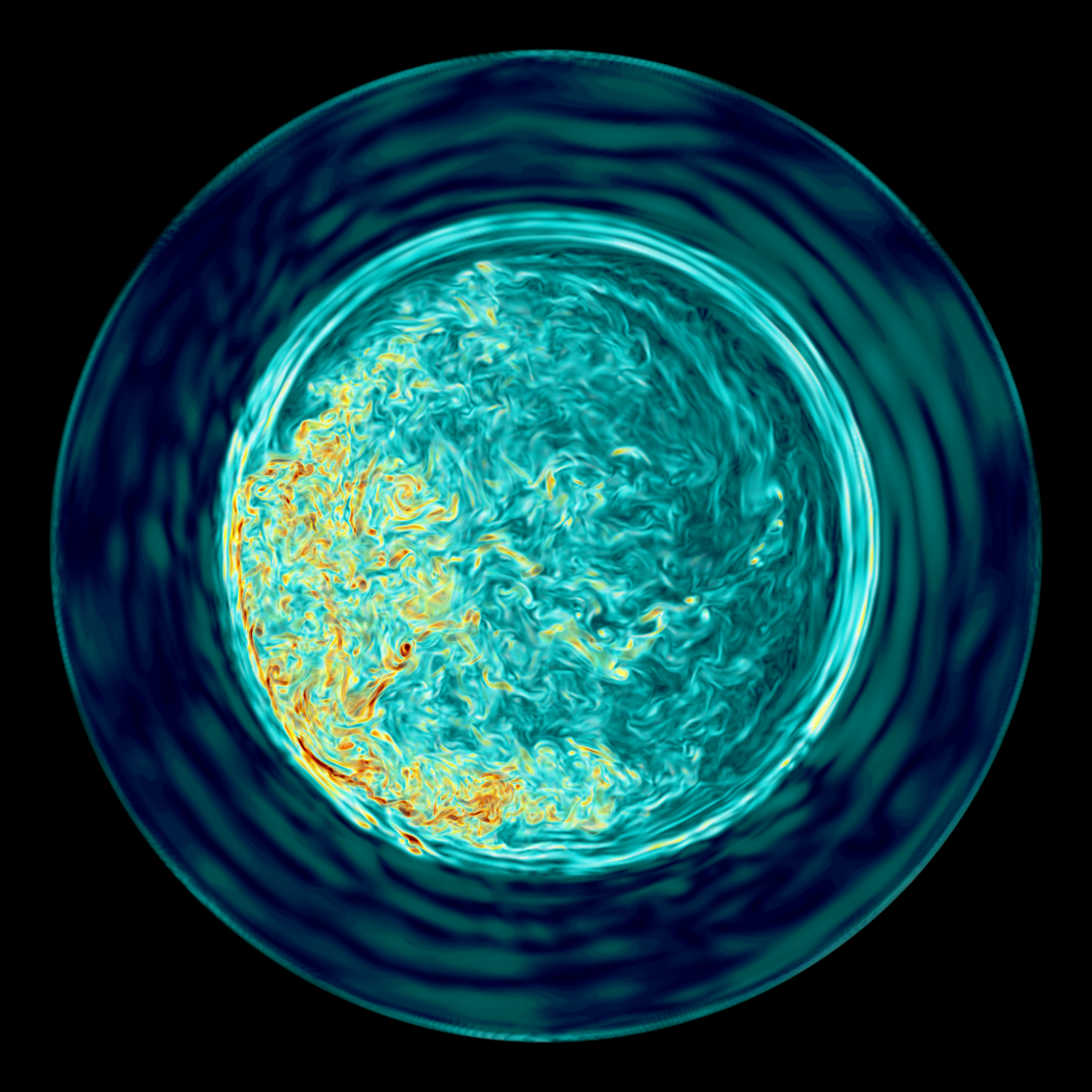}};
        \begin{scope}[x={(image.south east)},y={(image.north west)}]
            \draw[white,-Bar,-Triangle, line width=0.4mm] (0.5,0.95) -- (0.206,0.95) node[midway,above] {26 Mm};
            \draw[white,dashed] (0.206,0.95) -- (0.206,0.5);
            \draw[white,-Bar,-Triangle, line width=0.4mm] (0.5,0.05) -- (0.251,0.05) node[midway,below] {22 Mm};
            \draw[white,dashed] (0.251,0.05) -- (0.251,0.5);
        \end{scope}
    \end{tikzpicture}
    \caption{Center-plane slice rendering of the vorticity magnitude in run W32 (CPM setup, 768$^3$ grid, $\log L/L_{\star}=7$) at dump 1650 ($t=39.9\,$h). In the southwest quadrant, note the intrusion of the dipole circulation pattern of the convective core into the semiconvective layers. This is to be contrasted with the $\log L/L_{\star}=5$ simulation rendered in Figures~\ref{fig:W20bobs} and~\ref{fig:scvz-kw}, where the turbulent motions are confined to a spherical region circumscribed by a clear boundary at $R \simeq 22\,$Mm. White arrows indicate physical distances from the centre of the star.}
    \label{fig:W32-plume}
\end{figure}
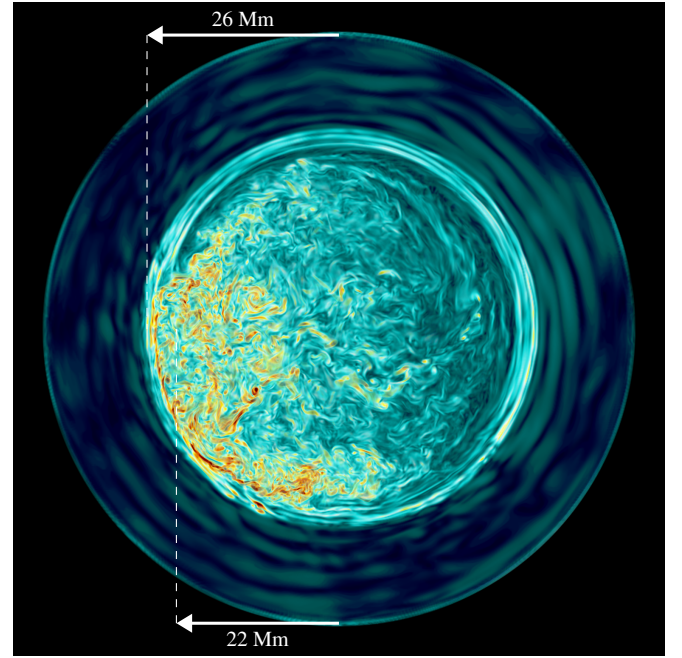

\begin{figure}
    \centering
	\includegraphics[width=\columnwidth]{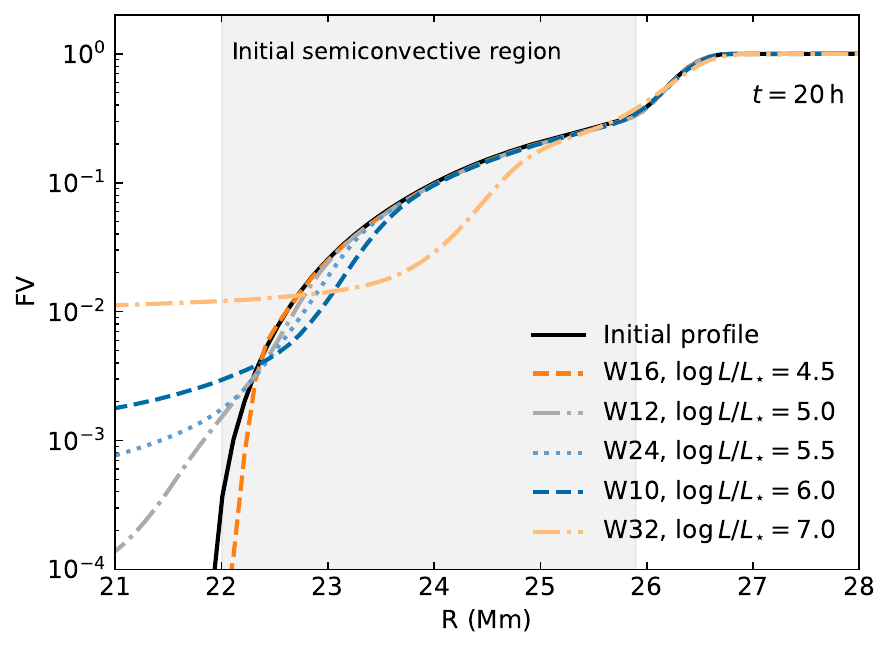}
    \caption{FV profile at $t=20\,$h for our CPM runs performed on a 768$^3$ Cartesian grid and with different heating rates (see legend). The black solid line corresponds to the initial setup. The semiconvection zone in the initial setup is shaded in grey. Note how the FV profile in the semiconvective layers is homogenized in the highest-heating run (W32).}
    \label{fig:heating_FV}
\end{figure}

We have seen in Figure~\ref{fig:diffusivity-IGW} that, while no mixing is measured throughout most of the semiconvection zone, $D>0$ in the layers close to the convective--semiconvective interface. This mixing is caused by strong convective motions that overshoot into the semiconvection zone. These overshooting motions are most clearly seen in our highest-heating runs, as exemplified by Figure~\ref{fig:W32-plume}. They gradually homogenize the semiconvection zone, thereby shrinking it to the benefit of the convective core. This is clearly shown in Figure~\ref{fig:heating_FV}. The black solid line shows the initial FV profile, and the coloured lines represent FV after 20\,h of star time for 768$^3$ simulations performed at different heating rates. The semiconvection zone is eroded to some extent in all runs, but this is most striking for the highest-heating run (W32). Since all fluid motions are faster at higher heating, the evolution is effectively fast-forwarded and it is therefore natural to observe a faster erosion of the semiconvection zone.

\begin{figure}
    \centering
	\includegraphics[width=\columnwidth]{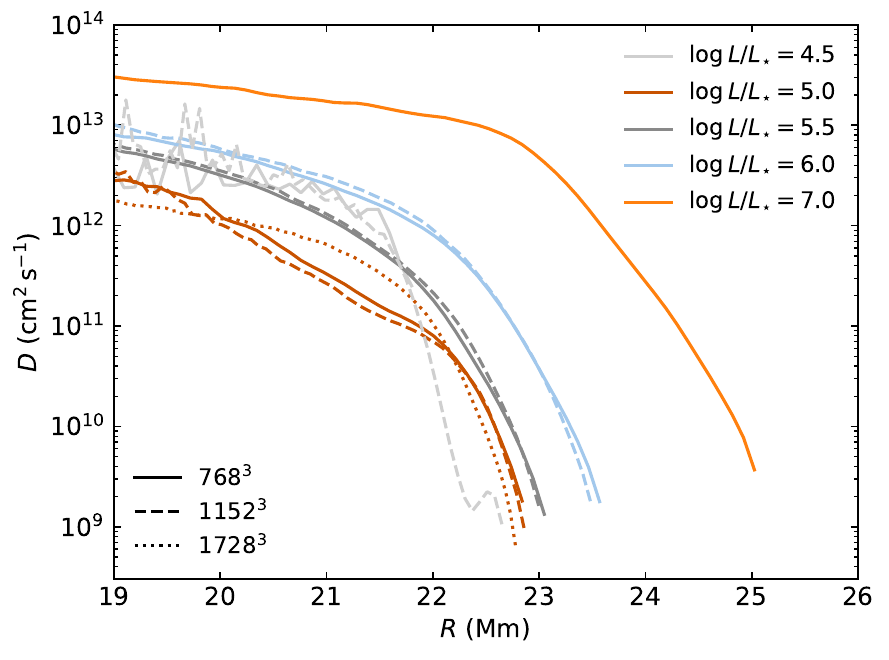}
    \caption{Measured diffusion coefficient in our CPM simulations (see legend for grid resolution and heating rate). For each run, $D$ was calculated by inverting the evolution of the FV profile between $t=20.0-22.5\,$h and $t=22.5-25.0\,$h.}
    \label{fig:diffusion-plumes}
\end{figure}

There is also evidence that the morphology of the flow changes at high heating rates, with the development of larger overshooting motions. When the heating rate is increased, the convective motions are able to overshoot further into the semiconvection zone, a phenomenon clearly seen in the movies available at \url{https://www.ppmstar.org}. This phenomenon can also be detected in diffusivity profiles. Figure~\ref{fig:diffusion-plumes} shows the measured diffusivity profiles in our CPM simulations, calculated in each case in the $t=20-25\,{\rm h}$ interval. Three important observations can be made. First, the agreement between simulations performed using different grid resolutions but identical heating rates is satisfactory. Second, as the heating rate is increased, the convective--semiconvective interface, whose location is marked by the radius at which $D$ abruptly drops off, is pushed further outward ($\simeq 22\,$Mm at $\log L/L_{\star}=5$ compared to $\simeq 23\,$Mm at $\log L/L_{\star}=7$). This is a direct consequence of the faster evolution of the high-heating simulations. Third, past the convective boundary, the slope of the diffusivity profile is steeper at low heating rates. All other things being equal, this signals a faster damping of the overshooting motions at low heating rates, consistent with our observation that overshooting motions intrude further into the semiconvection zone when the heating rate is increased.

Naturally, the astrophysically interesting question is to figure out whether these overshooting motions can homogenize the semiconvection zone at nominal heating. One way to attempt to answer this question is to establish a scaling relation for the core mass entrainment rate. Here, we define the entrained mass as being the total mass of envelope material located inward of $R=22\,$Mm. This definition is convenient as our \code{PPMstar} calculations follow the evolution of two fluids (a C/O-rich core fluid and an almost pure-He envelope fluid), but it will inevitably underestimate the true rate at which the convective core is eroding the semiconvective region in our simulations. Indeed, the semiconvective region is closer in composition to the convective core than to the stable envelope (${\rm FV}<0.5$), meaning that most of the entrained mass is not captured by our definition. As we will see, this issue has no impact on the main conclusion of the mass entrainment rate analysis.

\begin{figure}
    \centering
	\includegraphics[width=\columnwidth]{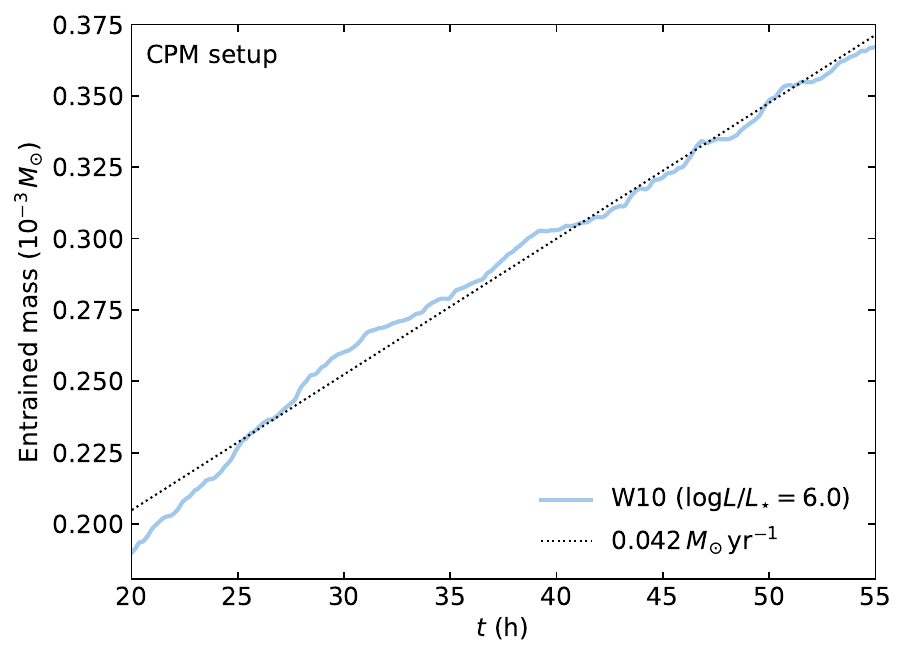}
	\includegraphics[width=\columnwidth]{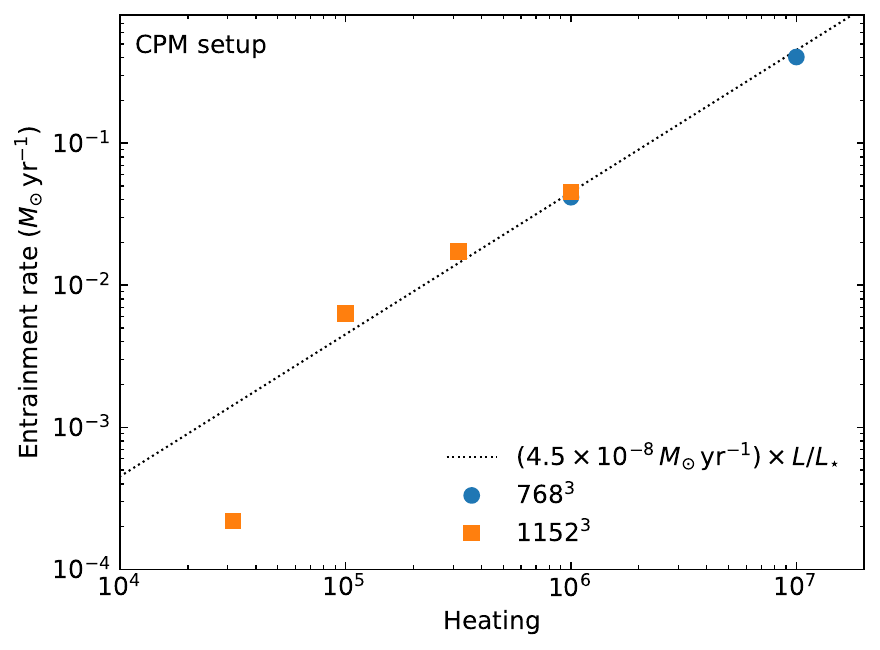}
    \caption{\textit{Top}: Time evolution of the entrained mass in run W10 (blue line, $\log L/L_{\star} =6$ and $768^3$ grid) and best-fit linear entrainment rate (black dotted line). \textit{Bottom}: Heating series of the measured mass entrainment rate (blue and orange symbols) and best-fit linear scaling law (black dotted line). The $\log\,L/L_{\star}=4.5$ run was omitted from the fit; its departure from the linear relation is likely a manifestation of the numerical issues that appear at low-Mach numbers with our explicit gas dynamics code (Section~\ref{sec:resolution}).}
    \label{fig:entrainment}
\end{figure}

The top panel of Figure~\ref{fig:entrainment} demonstrates how the mass entrainment rate is measured for a given simulation, and the bottom panel shows the mass entrainment scaling relation thus found. To stay clear from the initial transient, we have evaluated the mass entrainment rates using only $t >20\,$h, and for more robust measurements we only employed simulations longer than $40\,$h. As in previous works \citep{jones2017,andrassy2020,herwig2023,mao2023}, we find that the entrainment rate scales linearly with heating (if we exclude the lowest-heating run, consistent with our discussion in Section~\ref{sec:heating}). Extrapolating this relation to nominal luminosity yields an entrainment rate of $4.5 \times 10^{-8}\,M_{\odot}\,{\rm yr}^{-1}$. The 4\,Mm-thick semiconvection zone contains $0.08\,M_{\odot}$: assuming entrainment at this constant rate, it would take less than 2\,Myr to completely erase it. This is much shorter than any relevant evolutionary timescale. In particular, it is significantly faster than the rate at which the convective core grows in CHeB stars (a growth rate of $\simeq 3 \times 10^{-9}\,M_{\odot}\,{\rm yr}^{-1}$ can be inferred from Figure~\ref{fig:kip}). According to this analysis, we should therefore expect the semiconvection zone to be completely erased, implying that CHeB stars cannot have a semiconvection zone. In other words, a semiconvection zone cannot form in the first place, and the initial CPM setup is incorrect. Note that our restrictive definition of the entrained mass, which necessarily underestimates the total entrained mass in our simulations, can only strengthen this conclusion.

However, there is a major caveat with this analysis. As explained in Section~\ref{sec:evol}, the convective boundary may also be migrating for other spurious reasons related to the current \code{PPMstar} equation of state. When we measure an entrainment rate, we capture the sum of all the processes at play. We may therefore be overestimating the rate at which the semiconvection zone is eroded due to real mixing processes as opposed to the boundary migration due to the equation of state). That being said, independent results also indicate that the semiconvection zone should be quickly erased. \cite{anders2022b} have recently performed hydrodynamics simulations of an idealized plane-parallel setup containing a semiconvection zone between a convective and a stable region. They describe a similar behaviour to that detailed above, with overshooting convective motions gradually entraining low-$\mu$ material into the convective region and homogenizing the semiconvective layers. Ultimately, after thousands of convective turnover timescales, they report the disappearance of the semiconvection zone. 

\section{Conclusion}
\label{sec:conclu}
We have presented the first full-sphere 3D hydrodynamics simulations of the interior of a CHeB star. For dozens of convective turnover timescales, we followed the hydrodynamical response of our high-resolution simulations to two different initial stratifications (one with a semiconvective region and one without) for a 3\,$M_{\odot}$ CHeB star. 

We have recovered many of the key findings of our recent simulations of core convection in massive main-sequence stars, including the presence of a large dipole circulation pattern in the convective core, the excitation of a rich spectrum of IGWs in the stable envelope consistent with the eigenfrequencies predicted by \code{GYRE} for the same stratification, and the $L^{1/3}$ and $L^{2/3}$ scaling of the convective and IGW velocities, respectively. We found that the extended core of the PM prescription remains fully convective even if $\nabla_{\rm rad} - \nabla_{\rm ad}$ becomes very close to 0 (while remaining positive). We have described how an hypothetical semiconvection zone would be dominated by IGW motions. Despite the high amplitudes of these waves, we see no evidence for IGW mixing, in contrast with our recent finding for the core boundary region of massive main-sequence stars \citep{herwig2023}.

For simulations initialized with a semiconvective interface, we have also observed the incursion of convective motions inside the semiconvective zone, a phenomenon that gradually erases this region. While the efficiency of this mixing process at nominal luminosity remains unclear, it could be sufficient to homogenize the semiconvection zone of a CHeB star much more rapidly than any relevant evolutionary timescale. This would imply, as recently suggested by \cite{anders2022b}, that CHeB stars cannot harbour a semiconvective interface between their C/O-rich cores and He envelopes.

Future research should investigate other CBM prescriptions beyond the \code{MESA} CPM and PM schemes to model the CHeB phase. A growing body of work \citep{anders2022,andrassy2023,baraffe2023,blouin2023a,mao2023} is suggesting that the stratifications implied by these two CBM schemes is not appropriate. In particular, multi-dimensional hydrodynamics simulations consistently point to the formation of a convective penetration zone where the temperature gradient smoothly transitions from $\nabla_{\rm ad}$ to $\nabla_{\rm rad}$ over a fraction of a pressure scale height ($\sim 0.1-1\,H_P$). This is in clear tension with the discontinous transitions implied by \code{MESA}'s CPM and PM schemes.

\section*{Acknowledgements}
We thank the anonymous referee for their comments and suggestions, which have contributed to improving the clarity of this manuscript. SB is a Banting Postdoctoral Fellow and a CITA National Fellow, supported by the Natural Sciences and Engineering Research Council of Canada (NSERC). FH acknowledges funding through an NSERC Discovery Grant. PRW acknowledges funding through NSF grants 1814181 and 2032010. FH and PRW have been supported through NSF award PHY-1430152 (JINA Center for the Evolution of the Elements). The simulations presented in this work were carried out on the NSF Frontera supercomputer operated by the Texas Advanced Computing Center at the University of Texas at Austin and on the Compute Canada Niagara supercomputer operated by SciNet at the University of Toronto. The data analysis was carried on the Astrohub online virtual research environment (\url{https://astrohub.uvic.ca}) developed and operated by the Computational Stellar Astrophysics group (\url{http://csa.phys.uvic.ca}) at the University of Victoria and hosted on the Compute Canada Arbutus Cloud at the University of Victoria.

\section*{Data Availability}
Simulation outputs are available at \url{https://www.ppmstar.org} along with the Python notebooks that have been used to generate the figures presented in this work.

\bibliographystyle{mnras}
\bibliography{references}

\appendix
\section{Opacity model}
\label{sec:opacity}
The polynomial opacity model used in our \code{PPMstar} simulations has the form
\begin{equation}
\kappa = \sum_{i=0}^5 b_i \left( T_7 \right)^{5-i},
\end{equation}
where $T_7 = \log T - 7$.\footnote{Here and elsewhere, $\log$ stands for $\log_{10}$.} The $b_i$ coefficients are obtained via bilinear interpolation in the mass fraction and $\log R \equiv \log \rho - 3 \log T + 18$ space,
\begin{equation}
b_i = w_{11} a_{11}^i + w_{12} a_{12}^i +w_{21} a_{21}^i + w_{22} a_{22}^i.
\end{equation}
The $a_{jk}^i$ fit parameters are given in Table~\ref{tab:kappa_fit}, and the bilinear interpolation weights $w_{jk}$ are given by
\begin{equation}
\begin{split}
w_{11} &= (\log R_2 - \log R)(X_{{\rm CO},2} - X_{\rm CO})/\alpha, \\
w_{12} &= (\log R_2 - \log R)(X_{\rm CO} - X_{{\rm CO},1})/\alpha, \\
w_{21} &= (\log R - \log R_1)(X_{{\rm CO},2} - X_{\rm CO})/\alpha, \\
w_{22} &= (\log R - \log R_1)(X_{\rm CO} - X_{{\rm CO},1})/\alpha,
\end{split}
\end{equation}
with
\begin{equation}
\alpha = (\log R_2 - \log R_1)(X_{{\rm CO},2} - X_{{\rm CO},1}).
\end{equation}
$X_{\rm CO}$ is the combined mass fraction of C and O. We stress that this opacity model is reliable only within the limited composition--temperature--density space covered by our simulations. The chosen polynomial functional form of the fit is inherently flexible, specifically designed to allow for the same model (with different parameters) to be applicable across a wide range of stellar environments, as it is employed in \code{PPMstar} simulations of various types of stars \citep{blouin2023b,blouin2023a,mao2023}.

\begin{table}
\centering
\caption{Opacity model fit parameters}
\renewcommand{\arraystretch}{1.2} 
\begin{tabular}{lr|lr}
\hline
$\log R_1$ & $-$2.37164275 & $\log R_2$ & $-$1.77937633 \\
$X_{{\rm CO},1}$ & 0.00000000 & $X_{{\rm CO},2}$ & 0.73187004 \\
\(a_{11}^0\) & 0.29340722 & \(a_{12}^0\) & $-$0.59945332 \\
\(a_{11}^1\) & $-$1.02164622 & \(a_{12}^1\) & 2.72183141 \\
\(a_{11}^2\) & 1.13886313 & \(a_{12}^2\) & $-$5.12118290 \\
\(a_{11}^3\) & $-$0.27116766 & \(a_{12}^3\) & 5.12686032 \\
\(a_{11}^4\) & $-$0.39983029 & \(a_{12}^4\) & $-$2.99192563 \\
\(a_{11}^5\) & 0.44397946 & \(a_{12}^5\) & 1.08999269 \\
\(a_{21}^0\) & 0.30510341 & \(a_{22}^0\) & $-$0.51150380 \\
\(a_{21}^1\) & $-$1.01849526 & \(a_{22}^1\) & 2.94414203 \\
\(a_{21}^2\) & 0.97161241 & \(a_{22}^2\) & $-$7.04134649 \\
\(a_{21}^3\) & 0.13445119 & \(a_{22}^3\) & 8.72264878 \\
\(a_{21}^4\) & $-$0.88449144 & \(a_{22}^4\) & $-$5.95817987 \\
\(a_{21}^5\) & 0.69582918 & \(a_{22}^5\) & 2.15246120 \\
\hline
\end{tabular}
\label{tab:kappa_fit}
\end{table}

\bsp
\label{lastpage}

\end{document}